\documentclass[english,aps,nofootinbib,longbibliography,notitlepage,superscriptaddress,prd,preprintnumbers,showkeys, twocolumn]{revtex4-1}
\usepackage[latin9]{inputenc}
\usepackage{babel}
\usepackage{float}
\usepackage{graphicx}
\usepackage{hyperref}
\usepackage{amsmath}
\usepackage{amssymb}
\usepackage{todonotes}
\usepackage[mathscr]{eucal}
\usepackage{microtype}
\usepackage{esint}
\usepackage{amsbsy}
\usepackage{color}
\usepackage[capitalise]{cleveref}
\usepackage{breakurl}
\usepackage{ulem}
\usepackage{bbm}
\usepackage{mathtools}
\usepackage{braket}
\usepackage{comment}
\usepackage[super]{nth}
\usepackage{todonotes}
\usepackage{hyphenat}

\newcommand{\Mp}{M_\mathrm{Pl}}

\newcommand{\Meff}{M_\mathrm{eff}}
\newcommand{\dd}{\mathrm{d}}

\newcommand{\req}{r_{\text{eq}}}

\makeatletter

\pdfpageheight\paperheight
\pdfpagewidth\paperwidth

\@ifundefined{textcolor}{}{%
 \definecolor{BLACK}{gray}{0}
 \definecolor{WHITE}{gray}{1}
 \definecolor{RED}{rgb}{1,0,0}
 \definecolor{GREEN}{rgb}{0,1,0}
 \definecolor{BLUE}{rgb}{0,0,1}
 \definecolor{CYAN}{cmyk}{1,0,0,0}
 \definecolor{MAGENTA}{cmyk}{0,1,0,0}
 \definecolor{YELLOW}{cmyk}{0,0,1,0}
}

\makeatother

\makeatletter

\DeclareRobustCommand{\rcite}[1]{%
  \rcite@aux#1,\@nil{#1}%
}
\def\rcite@aux#1,#2\@nil#3{%
  \if\relax#2\relax
    Ref.~\cite{#3}%
  \else
    Refs.~\cite{#3}%
  \fi
}
\makeatother

\hypersetup{
    colorlinks = true,
    citecolor = {red},
    linkcolor = {blue},
    urlcolor = {blue},
}

\newcommand{\be}{\begin{equation}}
\newcommand{\ee}{\end{equation}}

\begin{document}

\title{Cosmological dynamics of multifield dark energy}

\author{Johannes R. Eskilt}
\email{j.r.eskilt@astro.uio.no}
\affiliation{Institute of Theoretical Astrophysics, University of Oslo, P.O. Box 1029 Blindern, N-0315 Oslo, Norway}

\author{Yashar Akrami}
\email{akrami@ens.fr}
\affiliation{CERCA/ISO, Department of Physics, Case Western Reserve University, 10900 Euclid Avenue, Cleveland, Ohio 44106, USA}
\affiliation{Laboratoire de Physique de l'\'Ecole Normale Sup\'erieure, ENS, Universit\'e PSL, CNRS, Sorbonne Universit\'e, Universit\'e de Paris, F-75005 Paris, France}
\affiliation{Observatoire de Paris, Universit\'e PSL, Sorbonne Universit\'e, LERMA, 75014 Paris, France}
\affiliation{Astrophysics Group \& Imperial Centre for Inference and Cosmology,
Department of Physics, Imperial College London, Blackett Laboratory,
Prince Consort Road, London SW7 2AZ, United Kingdom}

\author{Adam R. Solomon}
\email{soloma2@mcmaster.ca}
\affiliation{Department of Physics and Astronomy, McMaster University, Hamilton, Ontario L8S 4M1, Canada}
\affiliation{Perimeter Institute for Theoretical Physics, Waterloo, Ontario N2L 2Y5, Canada}

\author{Valeri Vardanyan}
\email{valeri.vardanyan@ipmu.jp}
\affiliation{Kavli Institute for the Physics and Mathematics of the Universe (WPI), UTIAS, The University of Tokyo, Chiba 277-8583, Japan}


\begin{abstract}
We numerically and analytically explore the background cosmological dynamics of multifield dark energy with highly nongeodesic or ``spinning" field-space trajectories. These extensions of standard single-field quintessence possess appealing theoretical features and observable differences from the cosmological standard model. At the level of the cosmological background, we perform a phase-space analysis and identify approximate attractors with late-time acceleration for a wide range of initial conditions. Focusing on two classes of field-space geometry, we derive bounds on parameter space by demanding viable late-time acceleration and the absence of gradient instabilities, as well as from the de Sitter swampland conjecture.
\end{abstract}

\preprint{}
\maketitle

\tableofcontents

\section{Introduction}
\label{sec:intro}

Finding the physical mechanism responsible for late-time cosmic acceleration \cite{SupernovaCosmologyProject:1998vns,SupernovaSearchTeam:1998fmf} is one of the most exciting challenges in cosmology. Phenomenologically the simplest possibility is a cosmological constant, which with cold dark matter constitutes the cosmological standard model, $\Lambda$CDM. Despite the remarkable empirical success of $\Lambda$CDM \cite{Planck:2018vyg}, an extensive effort has been dedicated to studying possible alternatives to the cosmological constant (see \rcite{Copeland:2006wr,Clifton:2011jh,Bamba:2012cp,Joyce:2014kja,Amendola:2015ksp,Koyama:2015vza,Bull:2015stt,Ishak:2018his,Ferreira:2019xrr,CANTATA:2021ktz} for reviews). This is largely motivated by the fine-tuning problem of the cosmological constant \cite{Martin:2012bt,Burgess:2013ara}. Even without such theory considerations, the wealth of current and upcoming precise cosmological probes is a strong driving force for testing the theoretically viable alternatives to the cosmological constant scenario.

Indeed, the discovery of any effects signaling deviations from the $\Lambda$CDM model is among the primary objectives of near-future major cosmological surveys, such as the \textit{Euclid} space mission \cite{Laureijs:2011gra,Amendola:2016saw}. The careful modeling and classification of theoretically possible signals are therefore of utmost interest and importance.

Dark energy beyond the cosmological constant is usually modeled with a single scalar field, with prominent examples being \textit{quintessence} \cite{Copeland:2006wr} and \textit{scalar-tensor gravity} \cite{Clifton:2011jh}. Quintessence is the simplest dynamical dark energy scenario, which assumes a minimally coupled canonical scalar field with a prespecified potential. Successful acceleration requires the potential to be nearly flat, with a mass scale of the order of the Hubble constant ($\sim 10^{-33}$~eV). As a result, the quintessence field acts as a smooth, unclustered energy component at observationally relevant scales \cite{Amendola:2015ksp}, and the primary prediction is a modified expansion history, uniquely determined by the shape of the potential.

While quintessence models are typically considered to have a single scalar field, low-energy effective theories arising from string theory generically predict the presence of multiple dynamical scalar fields \cite{Douglas:2006es,Arvanitaki:2009fg}. Multifield models have been thoroughly studied in the context of cosmic inflation \cite{Sasaki:1995aw,GrootNibbelink:2001qt,Lalak:2007vi,Ashoorioon:2009wa,Achucarro:2010jv,Achucarro:2010da,Achucarro:2012sm,Achucarro:2012yr,Pi:2012gf,Renaux-Petel:2015mga,Brown:2017osf,Achucarro:2017ing,Achucarro:2018vey,Aragam:2020uqi}. Although more scarcely, they have also been discussed in the context of dark energy; see, e.g., \rcite{Boyle:2001du,Kim:2005ne,vandeBruck:2009gp,Jimenez:2012iu,Vardanyan:2015oha,Leithes:2016xyh,Akrami:2017cir,Cicoli:2020cfj,Cicoli:2020noz,Akrami:2020zfz,Paliathanasis:2021fxi,Burgess:2021qti,Burgess:2021obw,Anguelova:2021jxu}. Building on our earlier work \cite{Akrami:2020zfz}, in this paper we provide a detailed study of multifield dark energy models featuring a nontrivial field-space geometry. Multifield dynamics is significantly richer compared to the single-field counterparts, and possesses several theoretically and phenomenologically appealing properties. Particularly, owing to the nontrivial geometry of the field space, a successful cosmic acceleration can take place even if the scalar potential is very steep. 

The latter is not only a conceptually novel possibility worth investigating further, but might also be of crucial importance in the context of possible theoretical restrictions of finding de Sitter--like solutions in quantum gravity, such as string theory. Particularly, it has been conjectured \cite{Obied:2018sgi,Agrawal:2018own,Garg:2018reu,Ooguri:2018wrx} that the effective low-energy descriptions of viable string theory embeddings impose restrictions on the relative slope of the scalar potentials, limiting them to be steep. While the validity of these \textit{swampland} conjectures is being actively debated \cite{Kachru:2003aw,Akrami:2018ylq,Cicoli:2018kdo,Kachru:2018aqn,Kallosh:2019axr}, they provide an additional motivation for multifield extensions of the standard quintessence scenario, and we will include them in our analyses for the sake of completeness.

One of the primary features of multifield scenarios is the possibility of strongly curved trajectories in field space. Particularly, our focus will be the ``spinning" regime where the fields rotate rapidly on almost-circular trajectories. Despite the highly dynamical nature of this regime, it predicts cosmological background solutions close to $\Lambda$CDM \cite{Akrami:2020zfz}. In this paper we investigate a comprehensive class of two-field dark energy models and address the question of how general the spinning regime is. We parametrize the field space with ``radial" and ``angular" fields $r$ and $\theta$, and we impose a $\theta$ shift symmetry on the field-space metric, $\mathcal{G}_{ab} = \mathrm{diag}\left( 1, f(r)\right)$. While in \rcite{Akrami:2020zfz} our focus was to provide a proof of concept based on a flat field space, $f(r) = r^2$, in the present analysis we study more general metrics. Our first choice is a generic power-law form for $f(r)$, and the second one is a hyperbolic metric with negative curvature.

While identifying any exact fixed points in cosmological evolution appears to be challenging, we numerically prove that spinning trajectories are generically achieved and maintained with a relatively simple choice for the two-field potential, and, importantly, for a wide pool of initial conditions. For all the cases we also explicitly demonstrate that the de Sitter swampland conjecture is satisfied for large regions in the parameter space. In addition to the aforementioned de Sitter conjecture, we also show that our models generically satisfy the distance conjecture \cite{Palti:2019pca} in the observationally relevant epochs.  

Even though the spinning solutions universally predict cosmological-constant--like cosmic backgrounds, the spinning multifield scenarios we discuss can be observationally distinguished from $\Lambda$CDM owing to their richer clustering properties. Particularly, the rapidly rotating multifield constructions possess a perturbation mode much heavier than the Hubble scale, leading to enhanced clustering on sub-Hubble scales. Moreover, the sound speed of the effective massless mode is significantly reduced, again, leading to enhanced subhorizon clustering of the dark energy component \cite{Akrami:2020zfz}.  

For each of the considered models we have identified a range of linear, subhorizon scales affected by the clustering, demonstrating the presence of a wide range of observationally accessible scales which are of interest for upcoming large-scale structure probes. The effective sound speed can turn imaginary in certain, though by far not all, cases, signaling the presence of gradient instabilities which limit the range of validity of the models. We provide summaries of observational viability, satisfaction of the de Sitter conjecture, and the absence of gradient instabilities, effectively limiting the parameter ranges of the considered models. One or more of these limits can be used as generic guiding principles when constructing multifield models.

\section{Multifield dark energy}
\label{sec:dark-energy-model}
In this paper, we consider a multifield model of dark energy where a number of scalar fields $\phi^a$ live in a field space with a nontrivial metric $\mathcal{G}_{ab}(\phi)$ and are minimally coupled to gravity, resulting in the action
\begin{equation}\label{eq:action}
	S\!=\!\!\int\!\!\dd^4x \sqrt{-g} \left[ \frac{\Mp^2}{2}R - \frac{1}{2}\mathcal{G}_{ab}(\phi) \partial_{\mu}\phi^a \partial^{\mu}\phi^b\! -\!V(\phi)\!+\!\mathcal{L}_\mathrm{m} \right].
\end{equation}
Here $R$ is the Ricci curvature for the spacetime metric $g_{\mu\nu}$, $\Mp$ is the Planck mass, $V(\phi)$ is the potential for the scalar fields, and $\mathcal{L}_\mathrm{m}$ is the matter Lagrangian.

\subsection{Background evolution and phase-space dynamics}
Considering the flat Friedmann-Lema\^itre-Robertson-Walker (FLRW) metric, the Friedmann equation is
\begin{equation}\label{eq:friedmann}
	3\Mp^2 H^2 = \frac{1}{2} \mathcal{G}_{ab} \dot{\phi}^a \dot{\phi}^b + V + \rho_\mathrm{m} \,,
\end{equation}
where $H \equiv \dot{a}/a$ is the Hubble expansion rate with $a$ the scale factor and $\rho_\mathrm{m}$ is the energy density of matter fields. An overdot denotes a derivative with respect to cosmic time $t$. The scalar field equations of motion are
\begin{equation}\label{eq:eom}
	D_t \dot{\phi}^a + 3H\dot{\phi}^a +V^a = 0\,,
\end{equation}
where $V_a \equiv \partial V / \partial \phi^a$ and $D_t$ is the field-space covariant time derivative defined as
\begin{equation}
	D_t A^a \equiv \dot{A}^a + \Gamma^a_{bc} A^b \dot{\phi}^c\,,
\end{equation}
with $\Gamma^a_{bc}$ the field-space Christoffel symbols.

Even though one can consider any number of scalar fields for the dark energy models we are interested in, in the rest of this paper we follow \rcite{Akrami:2020zfz} and restrict ourselves to two fields, $\phi^a=(r,\theta)$.\footnote{These should not be confused with physical radial or angular coordinates.} This suffices to display many of the novel features that arise when moving to multifield models. Consider a two-dimensional field-space metric of the form
\begin{equation}\label{eq:metric}
	\dd s^2_\mathrm{fields} = \dd r^2 + f(r, \theta)\dd\theta^2\,.
\end{equation}
It is often helpful to interpret $r$ and $\theta$ as polar coordinates in field space, hence the notation $\phi^a=(r,\theta)$ for the scalars. As we will see later, this is motivated by examples where the metric enjoys a shift symmetry in the $\theta$ direction, so that $f(r,\theta)=f(r)$. In this case \cref{eq:metric} is the most general field-space metric, up to field redefinitions.

For the metric \eqref{eq:metric}, the Friedmann equation \eqref{eq:friedmann} and the scalar field equations of motion \eqref{eq:eom} become
\begin{align}
    3\Mp^2H^2 = \frac{1}{2}\left(\dot{r}^2 + f \dot{\theta}^2 \right) + V + \rho_\mathrm{m}\,,\\
    \label{eq:r_eq_of_mot}
    \Ddot{r} + 3H\dot{r} + V_r - \frac{1}{2}f_r \dot{\theta}^2 = 0\,,\\
    \label{eq:theta_eq_of_mot}
    \Ddot{\theta} + 3H\dot{\theta} + \frac{1}{f}V_\theta + \frac{f_r}{f} \dot{r}\dot{\theta} + \frac{f_\theta}{2f}\dot{\theta}^2= 0\,,
\end{align}
where $V_r\equiv \partial V / \partial r$, $V_\theta\equiv \partial V / \partial \theta$, $f_r\equiv \partial f / \partial r$ and $f_\theta\equiv \partial f / \partial \theta$. In order to analyze the phase-space dynamics of these background equations, we rewrite them as a set of first-order differential equations,
\begin{widetext}
\begin{align}
	x^\prime_{r} &= 3x_{r}\left(x_{r}^2+x_{\theta}^2 - 1 + \frac{1+w_\mathrm{m}}{2} (1-x_{r}^2-x_{\theta}^2 - y^2)\right) +  \sqrt{3/2} k_1 x_{\theta}^2 - \sqrt{3/2} k_2 y^2\,,\label{x_r}\\
	x^\prime_{\theta} &= 3x_{\theta}\left(x_{r}^2 + x_{\theta}^2 - 1 + \frac{1+w_\mathrm{m}}{2}(1-x_{r}^2-x_{\theta}^2 - y^2) \right) -\sqrt{3/2} k_1x_{r}x_{\theta} - \sqrt{3/2}k_3y^2\,,\label{x_t}\\
	y^\prime &= 3y\left(x_{r}^2 + x_{\theta}^2 + \frac{1+w_\mathrm{m}}{2}(1-x_{r}^2-x_{\theta}^2 - y^2) \right) + \sqrt{3/2}y\left(k_2 x_{r} + k_3x_{\theta} \right)\,	\label{y},
\end{align}
\end{widetext}
where we have introduced the quantities
\begin{align}
	&x_{r} \equiv \frac{\dot{r}}{\sqrt{6}H \Mp}\,, \quad  x_{\theta} \equiv \frac{\sqrt{f}\dot{\theta}}{\sqrt{6}H \Mp}\,, \quad  y \equiv \frac{\sqrt{V}}{\sqrt{3}H \Mp}\,,\\
	&k_1 \equiv \Mp \frac{f_{r}}{f}\,, \quad k_2 \equiv \Mp \frac{V_{r}}{V}\,, \quad k_3 \equiv \Mp \frac{V_{\theta}}{\sqrt{f}V}\,,\label{eq:ks}
\end{align}
primes denote derivatives with respect to $N\equiv\ln a$,
and $w_\mathrm{m}$ is the equation of state for matter. Additionally, the Friedmann equation becomes the constraint
\begin{equation}\label{eq:Fried_const}
	1 = x^2_{r} + x^2_{\theta} + y^2 + \Omega_\mathrm{m}\,
\end{equation}
where $\Omega_\mathrm{m}$ is the fractional energy density parameter of matter fields. Note that \cref{x_r,x_t,y} do not directly depend on $f_\theta$, even though it appears in \cref{eq:theta_eq_of_mot}. We can also express the scalar fields' equation of state $w_\phi$ and fractional energy density parameter $\Omega_\phi\equiv\rho_\phi/(\rho_\phi+\rho_\mathrm{m})$, with $\rho_\phi$ the scalar fields' energy density, in terms of $x_r$, $x_\theta$, and $y$,
\begin{align}
    \label{eq:eq_of_st_x}
	w_{\phi} &= \frac{x_r^2 + x_{\theta}^2 - y^2}{x_r^2 + x_{\theta}^2 + y^2}\,,\\
    \label{eq:omega_x}
	\Omega_{\phi} &= x_r^2 + x_{\theta}^2 + y^2\,.
\end{align}

When $k_1$, $k_2$, and $k_3$ are constant, \cref{x_r,x_t,y} form a closed autonomous system, with some of its critical points possibly acting as attractors. If $k_3=0$, then the system is closed for $f=g(\theta)e^{\lambda r}$ and $V=ce^{\kappa r}$, with $\lambda$, $c$ and $\kappa$ constants, and $g(\theta)$ arbitrary. The critical points for $g(\theta)=1$ have been studied in \rcite{Cicoli:2020cfj,Cicoli:2020noz}. We are interested in cases where the potential depends on $\theta$ (in order to obtain spinning solutions \cite{Akrami:2020zfz}) and therefore $k_3\neq0$. In this case, there is no choice of $f$ and $V$ which will set $k_{1,2,3}$ constant simultaneously: requiring $k_{1,2}$ constant sets both $f$ and $V$ to be of the form $g(\theta)e^{\lambda r}$, in which case $k_3$ necessarily has $r$ dependence going as $f^{-1/2}$. Note, however, that the choice $f \sim g(\theta)^2$ and $V \sim e^{\lambda r+g(\theta)}$ with $g(\theta) \sim e^{\beta \theta}$ leads to constant $k_{1, 2, 3}$. We do not consider this case since it leads to the function $f$ being independent of $r$. Since at least one of the $k_i$ always depends on one of the fields, \cref{x_r,x_t,y} do not represent a closed system of equations, and identification of exact attractor points is challenging. Later we will prove numerically and via approximate treatment of the dynamical system that cosmologically relevant approximate (slowly varying in time) attractor points do in fact exist.

\subsection{Swampland conjectures}
\label{sec:swampland}

It is shown in \rcite{Akrami:2020zfz} that the scalar fields of models \eqref{eq:action} can maintain their motion on a steep potential while leading to a small {\it dark energy slow-roll parameter}
\begin{equation}
    \epsilon_\phi = \frac{3}{2}(w_\phi + 1)=\frac{3}{2}\frac{\dot\phi^2}{\frac{1}{2}\dot\phi^2+V}\,,
\end{equation}
where
\begin{align}
    \dot\phi^2 \equiv \mathcal{G}_{ab}\dot\phi^a\dot\phi^b\,.\label{eq:phidotsqr}
\end{align}
An $\epsilon_\phi \ll 1$ is required for the Universe to be accelerating at late times, as $\epsilon_\phi$ approaches a small {\it Hubble slow-roll parameter} $\epsilon \equiv - \dot{H}/H^2$ when dark energy dominates. Additionally, the scalar fields move in field space with a turning rate
\begin{equation}\label{eq:OmegaGen}
    \Omega = |D_t \mathcal{T}|\,,
\end{equation}
where $\mathcal{T}$ is the normalized tangent vector to the field-space trajectory given by
\begin{equation}\label{eq:tangent}
    \mathcal{T}^a=\frac{\dot\phi^a}{\dot\phi}\,.
\end{equation}
A nonzero $\Omega$ signals that the scalars move along a trajectory in field space that is not a geodesic, which is a hallmark feature of multifield dynamics. The turning rate \eqref{eq:OmegaGen} can be written as
\begin{align}
    \Omega = \frac{1}{\sqrt{|\det\mathcal{G}|}}\frac{|\dot\phi_1V_2 - \dot\phi_2V_1|}{\dot{\phi}^2}\, \label{eq:Omega2D}
\end{align}
for models with two scalar fields $\phi^1$ and $\phi^2$, where $\dot\phi_a=\mathcal{G}_{ab}\dot\phi^b$. 
The dark energy slow-roll parameter $\epsilon_\phi$ is then related to the so-called {\it potential slow-roll parameter}
\begin{equation}
    \epsilon_V\equiv\frac{\Mp^2}{2}\frac{\mathcal{G}^{ab} V_a V_b}{V^2}\,,
\end{equation}
through the relation
\begin{align}
    \epsilon_\phi = \epsilon_V\Omega_\phi\left(1 + \frac{\Omega^2}{9H^2}\right)^{-1}\,.
\end{align}
Given that the turning rate can be arbitrarily large in these dark energy models, i.e., $\Omega\gg H$, it is possible for $\epsilon_\phi$ to be small for arbitrarily large $\epsilon_V$, i.e., for arbitrarily steep potentials. This means that the condition required by the swampland {\it de Sitter conjecture},
\begin{equation}
    \label{eq:de_sitter_conjecture}
	\frac{|\nabla V|}{V}=\frac{\sqrt{\mathcal{G}^{ab} V_a V_b}}{V} \geq \frac{c}{\Mp}\,,
\end{equation}
with $c$ a constant of $\mathcal{O}(1)$, can be satisfied in these models.

In addition to the de Sitter conjecture, we can also consider the swampland {\it distance conjecture}. This is the conjecture that when the fields traverse a Planckian distance in field space, an infinite tower of light modes (e.g., Kaluza-Klein modes) appears, spoiling the effective field theory. More concretely, the ``refined" distance conjecture states that these light modes appear with a mass scale \cite{Palti:2019pca}
\begin{equation}
\label{eq:distance_conjecture}
	 M\sim M_\mathrm{ini} e^{-\lambda \Delta \phi / \Mp}\,,
\end{equation}
where $\lambda  = \mathcal{O}(1)$, $M_\mathrm{ini}$ is some initial mass scale, and $\Delta \phi$ is the field-space distance traveled by the scalar fields. This then means that the effective field theory breaks down when
\begin{equation}
	\Delta \phi \equiv \int\dd t\sqrt{\mathcal{G}_{ab}\dot\phi^a\dot\phi^b} \gtrsim \lambda^{-1} \Mp.\label{eq:distance1}
\end{equation}
Using \cref{eq:eq_of_st_x,eq:omega_x}, it is easy to show that $\Delta \phi$ can be expressed as
\begin{align}\label{eq:distance2}
	\nonumber
	\Delta \phi &= \sqrt{6}\Mp \int \dd N \sqrt{x_r^2 + x_{\theta}^2} \\
	&= \sqrt{3}\Mp \int \dd N \sqrt{(1+w_{\phi})\Omega_{\phi}}\,.
\end{align}
The second equation in \eqref{eq:distance2}, which can be proved to be independent of both the number of fields and the field-space metric, clearly shows that if the fields move forever and never settle at a local minimum, then the field-space distance will diverge. The Universe will eventually reach a point where a tower of light states is expected to appear due to the distance conjecture, resulting in a breakdown of the effective field theory. We will, however, show that for the class of models we study in this paper this breakdown will not happen in the near future. It is also interesting to note that \cref{eq:distance2} is analogous to what happens during inflation where the field-space distance can be written as a function of the tensor-to-scalar ratio $r$,
\begin{equation}
    \Delta \phi = \Delta N \sqrt{\frac{r}{8}}\Mp\,,
\end{equation}
with $\Delta N$ the number of inflationary $e$-folds.

In the rest of this paper, we will investigate to which degree both the de Sitter and the distance conjectures are satisfied for specific classes of multifield dark energy models, in addition to studying their cosmological dynamics. While different string-theory-based models predict different explicit values for the constants $c$ and $\lambda$ in \eqref{eq:de_sitter_conjecture} and \eqref{eq:distance_conjecture}, here we do not make any assumptions about their values and only assume that both are of $\mathcal{O}(1)$.

\section{Explicit models}
\label{sec:explicit-model}

We now perform analytical and numerical analyses of specific and simple classes of multifield dark energy models, and show explicitly how easily large turning rates can be achieved without tuning the initial conditions of the dynamical system. We also demonstrate that such models generically satisfy theoretical viability conditions imposed by the swampland conjectures.

We consider a potential of the form
\begin{equation}\label{eq:potential}
	V(r,\theta)= V_0-\alpha \theta + \frac{1}{2}m^2(r-r_0)^2\,,
\end{equation}
where $V_0$, $\alpha$, $m$ and $r_0$ are free parameters to be determined observationally. As argued in \rcite{Akrami:2020zfz}, this is a minimal potential in an effective field theory framework which allows for the ``spinning" solutions we wish to study, consisting of a mass term (after shifting $r$) and a term softly breaking the $\theta$ shift symmetry.\footnote{This is not an exhaustive study of all two-field models; see \rcite{Cicoli:2020cfj,Cicoli:2020noz,Anguelova:2021jxu} for other examples of cosmologically viable choices of $f$ and $V$.} Additionally, we assume that the field-space metric \eqref{eq:metric} only depends on $r$ and consider two specific forms for the field-space function $f(r,\theta)=f(r)$:\footnote{From now on, we set $\Mp = 1$.}
\begin{itemize}
	\item {\bf Power-law metric with} {\boldmath $f(r) = r^p$}: The Ricci curvature corresponding to this field-space metric is $\mathcal{R} = -p(p-2)/(2r^2)$, and we allow the parameter $p$ to take any positive or negative values. This means that, depending on the value of $p$, the field space can be positively curved [for $p\in (0,2)$], flat (for $p=0,2$), or negatively curved (for $p<0$ and $p>2$). For the flat metric with $p=2$, the fields $r$ and $\theta$ have a direct polar coordinate interpretation. This is the case studied in \rcite{Akrami:2020zfz}, where it was shown that the de Sitter condition is satisfied with a dark energy equation of state $w_{\phi} \approx -1$.
	
	The power-law metric $f(r) = r^p$ leads to singularities in the $\theta$ equation of motion \eqref{eq:theta_eq_of_mot} for $r=0$; this is due to the $1/f$ factors appearing in the terms on the left-hand side of the equation. In our studies of the power-law metric in this paper, we assume that the $r$-field vacuum expectation value (VEV), i.e., $r_0$ in the potential \eqref{eq:potential}, is nonzero, and we restrict ourselves to $r_0 > 0$. As we will see later, $r$ will then remain nonzero for the cosmological solutions of interest, avoiding singularities in \cref{eq:theta_eq_of_mot}. Unless otherwise stated, we set $r_0$ to $7\cdot10^{-4}$, but we find that the evolution of dark energy fields for the power-law metric $f(r) = r^p$ and the potential \eqref{eq:potential} is insensitive to the value of $r_0$ when $r_0 \lesssim 10^{-3}$. This is because, as we will see, at late times $r$ reaches a semiequilibrium value which is much larger than $r_0$.
	\item {\bf Hyperbolic metric with} {\boldmath $f(r) = e^{\beta r}$}: The Ricci curvature for this metric is $\mathcal{R} = -\beta^2 / 2$ with $\beta$ constant. This means that the field space is negatively curved for all nonzero values of $\beta$; $\beta=0$ corresponds to a flat field space. Unlike the power-law field-space metric, here $r=0$ does not lead to any singularities in the equations of motion and we therefore choose to set $r_0 = 0$. We find that different values of $r_0$ result in negligible differences in the evolution of dark energy fields, as long as $r_0 \lesssim 10^{-3}$.
	
	The hyperbolic metric $f(r) = e^{\beta r}$ corresponds to hyperbolic space $H_2$ or, equivalently, Euclidean AdS$_2$ with radius $2/\beta$.
	The scalar fields parametrize an SL(2,R)/SO(2) coset space. Consequently, the field space is invariant under M\"obius transformations acting on $\tau\equiv \frac\beta2\theta + i e^{-\beta r/2}$,
\begin{equation}
    \tau \to \frac{a\tau + b}{c\tau + d}\,,
\end{equation}
with $a,b,c,d$ real constants satisfying $ad-bc=1$. This is a three-parameter group, including the $\theta$ shift symmetry ($a=d=1$, $c=0$), a scaling symmetry ($b=c=0$, $ad=1$), and an SO(2) subgroup ($a=d=\cos\varphi$, $c=-b=\sin\varphi$). The presence of a potential generically breaks these, though can preserve a subgroup. Dark energy with a hyperbolic field space has also been studied in \rcite{Burgess:2021obw}.

\end{itemize}

\begin{figure}
	\centering
	\includegraphics[width=1.0\linewidth]{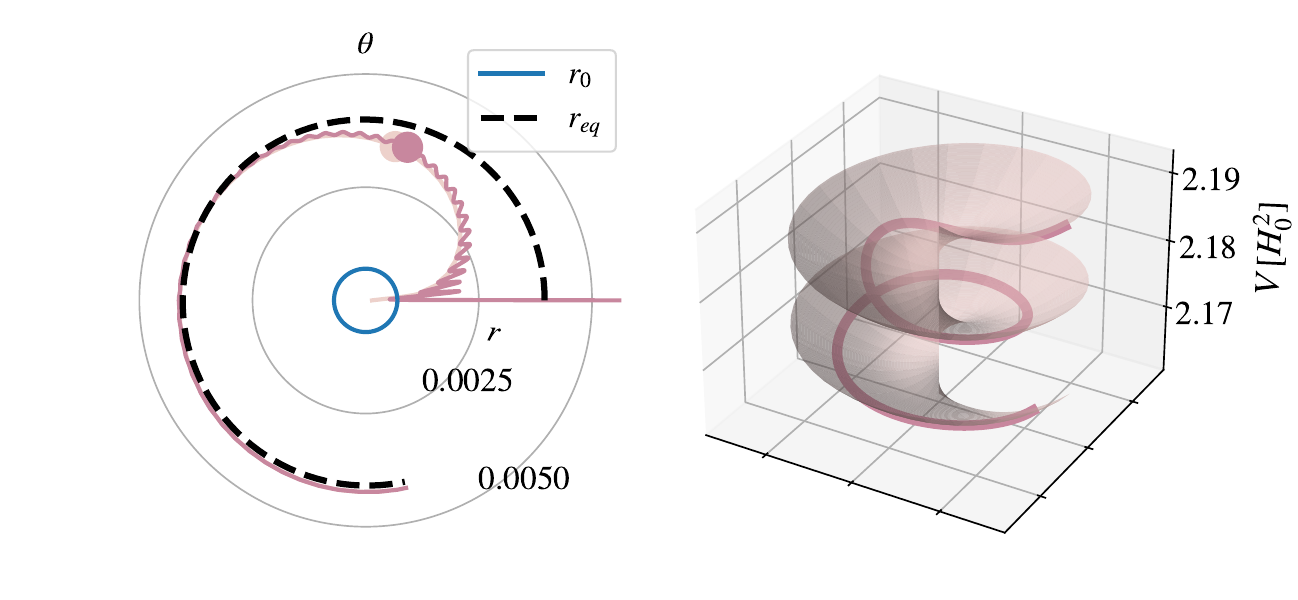}
	\caption{Examples of polar-coordinate  evolution of dark energy fields $r$ and $\theta$. Here, we have assumed a power-law field-space metric with $p=2$ and a potential of the form \eqref{eq:potential}. The two trajectories shown in the left plot correspond to two different (and arbitrary) initial conditions $r_\mathrm{ini} = 0.2 r_0$ and $r_\mathrm{ini} = 8r_0$, both of which converge toward the analytic semiequilibrium value $\req$ given by \cref{eq:p_metric_req}. The colored dots indicate where we are today. The right plot shows the trajectory of the fields as they roll down the potential. For better visualization in the left plot, we have divided the $\theta$ field by $30$. We have set $V_0=2.19H_0^2$, $\alpha = 2\cdot10^{-3}H_0^2$, $m=50H_0$, and $r_0=7\cdot 10^{-4}$ in both plots.}
	\label{fig:polar}
\end{figure}

\subsection{Power-law field-space metric}\label{sec:pol}

We start our analysis of the explicit models specified by the potential \eqref{eq:potential} by first analyzing cosmological solutions for the power-law field-space metric, $f(r)=r^p$. In order to understand the cosmological dynamics of these models, we provide in \cref{fig:polar} the polar-coordinate representation of a typical evolution of the scalar fields $r$ and $\theta$. We have set $p=2$, which is the specific case of power-law-metric models studied in \rcite{Akrami:2020zfz}. In the left panel of the figure, we present the evolution of the fields with time for two different $r$-field initial conditions---we have set the initial value of $\theta$ to zero for simplicity and without loss of generality.

In the early moments of cosmic evolution, and as the Hubble friction decreases, the $r$ field begins to move. The figure shows that, independent of its initial value, $r$ climbs up the potential until it reaches a semiequilibrium value $\req > r_0$, which varies very slowly with time. The figure also shows that the $\theta$ field increases with time, which is expected by the fact that the potential contains a $-\alpha \theta$ term.

Although the exact evolution of the fields is model dependent and can only be obtained numerically, it is possible to find an analytical approximation to the semiequilibrium value $\req$ by using \cref{x_r,x_t,y} and assuming that dark energy has almost fully dominated the matter fields, i.e., $\Omega_{\phi} \approx 1$. By assuming $r\approx\req$, or equivalently $x_r \approx 0$, \cref{x_r,y} then lead to two simple equations for $x_{\theta}$,
\begin{align}
    \label{eq:x_theta}
	x_{\theta} &\approx \sqrt{\frac{k_2}{k_1+k_2}}\,, \\
	\label{eq:x_theta_2}
	x_{\theta} &\approx -\frac{1}{\sqrt{6}} k_3\,.
\end{align}
Here, we have additionally assumed $x^\prime_r\approx0$, $y^\prime\approx 0$, $w_\mathrm{m}\approx0$ (as matter fields are nonrelativistic at late times and in the future) and $\Omega_\mathrm{m}\approx 0$ (as dark energy is fully dominant); the latter in combination with \cref{eq:Fried_const} implies $y^2\approx 1-x_\theta^2$, which we have used in deriving \cref{eq:x_theta,eq:x_theta_2}. These assumptions are valid for a large class of interesting cosmological solutions as they are equivalent to the justified assumption that the kinetic contributions to the total energy of dark energy fields are much smaller than the contribution from the potential, i.e., $x_{r} \ll y$ and $x_{\theta} \ll y$.

Equation \eqref{eq:x_theta} can also be written as
\begin{equation}
    \label{eq:x_theta2}
	x_{\theta}\approx \sqrt{\frac{k_2}{k_1}}y\,,
\end{equation}
which means that the conditions $k_2 \ll k_1$ and $y\approx 1$ are equivalent. These approximations are then valid as long as $x_\theta\ll 1$, in which case \cref{eq:eq_of_st_x} implies that $w_\phi\approx -1$. Without assuming $k_2 \ll k_1$, and therefore neglecting $k_2$ in the denominator of \cref{eq:x_theta}, the equation for $\req$ would be cumbersome. Here, we therefore restrict ourselves to cosmological solutions with $w_\phi\approx -1$ and assume
\begin{equation}\label{eq:x_theta3}
x_{\theta}\approx \sqrt{k_2/k_1}\,.
\end{equation}
We will check the validity of this approximation for the parameter choices we make in obtaining our numerical solutions and will comment on cases where the approximation does not hold.

By combining \cref{eq:x_theta_2,eq:x_theta3} and using \cref{eq:ks} for $k_1$, $k_2$, and $k_3$, we obtain the approximate equation
\begin{equation}
  \label{eq:general_req}
	(\req-r_0)^3 + \frac{2}{m^2}\left(V_0 - \alpha \theta \right)(\req-r_0) =  \frac{\alpha^2}{3m^4}\frac{f_r(\req)}{f(\req)^2}\,
\end{equation}
for a general field-space metric $f(r)$ and the potential \eqref{eq:potential}. In addition to providing an analytical way to find an approximate value for the semiequilibrium quantity $\req$, \cref{eq:general_req} also helps us better understand the cosmological dynamics and important features of our models.

Let us start with the turning rate $\Omega$ given by \cref{eq:OmegaGen,eq:Omega2D}. Assuming $r$ is almost constant, i.e., $r\approx\req$, we can neglect the $\dot r$ and $\ddot r$ terms in \cref{eq:Omega2D,eq:r_eq_of_mot}, the combination of which then gives
\begin{equation}
	\label{eq:turningrate_general}
	\Omega^2 = \frac{f_r(\req)}{2f(\req)}V_r(\req) = \frac{f_r(\req)}{2f(\req)}m^2 (\req-r_0).
\end{equation}
Here, we have assumed that $f(r)$ is positive, which is necessary for avoiding ghost instabilities, as $f(r)$ multiplies one of the kinetic terms in the action \eqref{eq:action}. Since $\Omega^2$ is in general either zero or positive, we must require $(\req-r_0) f_r(\req)\geq0$. As a nonzero (and sufficiently large) turning rate $\Omega$ (for which $\Omega/H \gg 1$) is required for the turning solutions we are seeking, $\req$ must be different than $r_0$, so the field $r$ is pushed away from its VEV, $r_0$. There is clearly a large family of $f(r)$ which provide nonzero values for $\req-r_0$. Additionally, \cref{eq:general_req} tells us that in order for any real, nonzero solution $\req-r_0$ to exist, $f_r(\req)$ has to have the same sign as $\req - r_0$. This means that the requirement $(\req-r_0) f_r(\req) > 0$ (for $\Omega\neq0)$ is always satisfied, independently of the form of $f(r)$.

Focusing now on the power-law metric, $f(r)= r^p$, and assuming that $\req \gg r_0$, \cref{eq:general_req} becomes
\begin{equation}
    \label{eq:app_req_1}
	\req^{p+4} + \frac{2(V_0-\alpha \theta)}{m^2}\req^{p+2} = \frac{p \alpha^2}{3m^4}\,.
\end{equation}
This equation cannot be solved for a general $p$, but we can find an approximate analytic expression for $\req$ in certain cases. Let us write \cref{eq:app_req_1} as
\begin{equation}
    \label{eq:app_req_2}
	\req^{p+2}\left(\req^2 + \frac{2(V_0-\alpha \theta)}{m^2}\right) = \frac{p \alpha^2}{3m^4}\,,
\end{equation}
which immediately tells us that if $\req^2 \ll 2(V_0-\alpha \theta)/m^2$, then
\begin{equation}
    \label{eq:p_metric_req}
	\req \approx \left(\frac{p \alpha^2}{6m^2(V_0-\alpha\theta)} \right)^{\frac{1}{p+2}}\,.
\end{equation}
The formula \eqref{eq:p_metric_req} therefore provides a good approximation for the semiequilibrium value $\req$ if
\begin{equation}
    \label{eq:p_metric_req_cond}
	\left(\frac{p \alpha^2}{6m^2(V_0-\alpha\theta)} \right)^{\frac{2}{p+2}}\ll\frac{2(V_0-\alpha\theta)}{m^2}\,.
\end{equation}
For the example we have presented in \cref{fig:polar}, we have shown both the exact numerical solution and the approximate semiequilibrium value $\req$ calculated using \cref{eq:p_metric_req}. The figure shows excellent agreement between the numerical solution and $\req$ for the two representative $r$-field initial conditions, and we have checked that the condition \eqref{eq:p_metric_req_cond} is satisfied for the set of parameter values we have chosen for that example. In general, our numerical analysis shows that $\alpha \theta \ll V_0$ over the entire cosmic history for a large set of parameter values, and therefore, if the parameters of a model satisfy the condition \eqref{eq:p_metric_req_cond} for $\theta=0$ (i.e., the initial value of $\theta$), then the condition is also satisfied at later times. Since $\theta$ increases with time, $V_0$ and $\alpha\theta$ eventually become comparable, but this does not happen in the past or even in the near future for the set of parameter values considered in this paper. One should, however, note that even though the condition \eqref{eq:p_metric_req_cond} and, consequently, the expression \eqref{eq:p_metric_req} are valid for the entire history, $\req$ slowly increases with time as the quantity $V_0-\alpha\theta$ in the denominator of \cref{eq:p_metric_req} decreases. We will see this slight change in the value of $\req$ for some of the cases we study later in this paper. Finally, the expression \eqref{eq:p_metric_req} also shows that in order to have a positive $\req$, $p$ must be positive since it appears in the numerator of the expression.

The other interesting quantity to study is the dark energy equation of state $w_\phi$. By combining \cref{eq:eq_of_st_x} and the approximate equation \eqref{eq:x_theta2}, we obtain
\begin{equation}
    \label{eq:general_eq_of_state}
    w_{\phi} = \frac{k_2-k_1}{k_2+k_1}= -1 + 2\left(1+\frac{f_r}{f}\frac{V}{V_r}\right)^{-1}\,,
\end{equation}
where we have also used eqs. \eqref{eq:ks} for $k_1$ and $k_2$ in terms of the field-space metric function $f(r)$, the potential $V$, and their derivatives with respect to $r$. We know that today $w_{\phi}$ is close to $-1$, and we expect it to stay close to $-1$ as dark energy fully dominates, i.e., as $\Omega_{\phi} \rightarrow 1$. By requiring the equation of state to be close to $-1$ when the $r$ field is almost at its equilibrium value $\req$, \cref{eq:general_eq_of_state} leads to the condition
\begin{equation}
     \frac{f_r(\req)}{f(\req)}\left(\frac{V_0 -\alpha \theta}{m^2 (\req-r_0)} + \frac{1}{2}(\req-r_0) \right) \gg 1
\end{equation}
for the potential \eqref{eq:potential}.
Rewriting this in terms of the turning rate of \cref{eq:turningrate_general}, we obtain
\begin{equation}\label{eq:ineq_Omega2}
   \left(\frac{f_r(\req)}{f(\req)} \right)^2 \frac{V_0 -\alpha \theta}{2\Omega^2} + \frac{\Omega^2}{m^2} \gg 1\,,
\end{equation}
which means that at least one of the two terms on the left-hand side of the inequality must be large. For the power-law metric with $f(r)=r^p$ and by taking into account that $\req\gg r_0$ for the solutions we consider in our analysis,  \cref{eq:turningrate_general} implies that
\begin{equation}
    \frac{\Omega^2}{m^2}\approx \frac{p}{2}\,.
\end{equation}
This means that for the values of $p$ that are not much larger than $2$, which are the cases we consider in this paper, the quantity $\Omega^2/m^2$ cannot be much larger than unity. The inequality \eqref{eq:ineq_Omega2} then tells us that 
\begin{equation}\label{eq:ineq_Omega2_pol}
   \frac{p^2}{\req^2} \frac{V_0 -\alpha \theta}{2\Omega^2} \gg 1\,.
\end{equation}

Given that $V_0$ is of $\mathcal{O}(H_0^2)$, $\alpha\theta \ll V_0$, and $\Omega\gg H$,\footnote{Note that this condition of $\Omega$ being large is required for steep potentials to provide viable late-time cosmic acceleration and consequently satisfy the de Sitter swampland condition (see \cref{sec:swampland,sec:swamp_explicit}); otherwise, there are choices of parameters for each model with which the potential is shallow, $\Omega$ is small, and the de Sitter condition is consequently not satisfied. In these cases, the models reduce, effectively, to single-field quintessence and can still provide viable cosmic acceleration.} we obtain the hierarchical inequality
\begin{equation}
    \label{eq:ineq_req_pol}
   \frac{1}{\req^2} \gg \frac{\Omega^2}{H_0^2} \gg \left(\frac{H}{H_0} \right)^2\,
\end{equation}
for the semiequilibrium value $\req$ and the turning rate $\Omega$.

\begin{figure}
	\centering
	\includegraphics[width=1\linewidth]{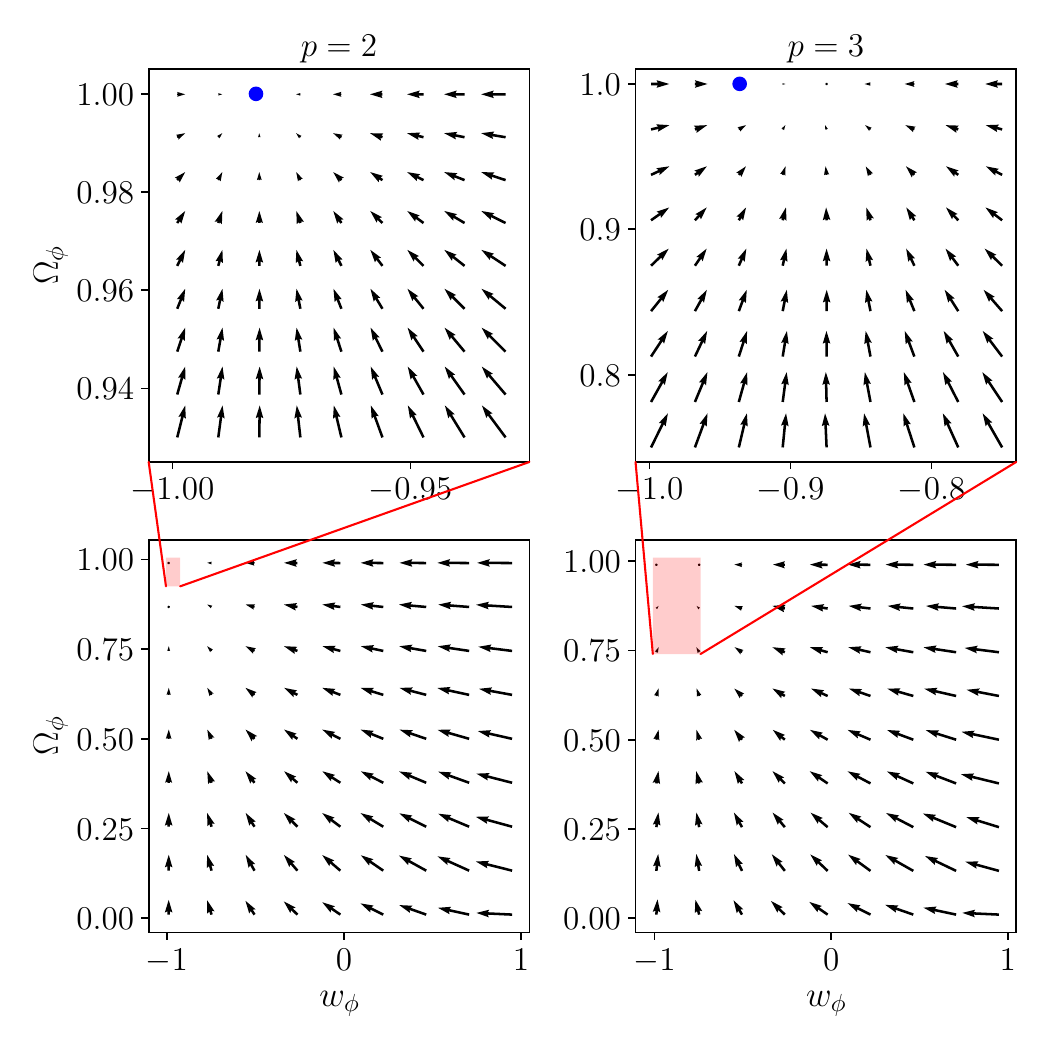}
	\caption{Phase-space diagrams for the evolution of dark energy fields in terms of the dark energy fractional density parameter $\Omega_{\phi}$ and equation of state $w_{\phi}$ for the power-law field-space metric with $p=2$ and $p=3$. Here, we have obtained the initial conditions for $x_{\theta}$ and $y$ from $\Omega_{\phi}$ and $w_{\phi}$, while we have set $x_r=0$ initially. The arrows indicate in which directions the fields have traveled during one $e$-fold and not the instant directions of motion. They therefore show the averaged behavior of the fields rather than the (oscillating) instantaneous behavior. Each upper panel is the zoomed version of the red box in the corresponding lower panel, and the blue dots show the approximate analytic solutions given by the combination of \cref{eq:w_eq_p_any,eq:p_metric_req} with $\theta=0$. We have set $V_0=2.19H_0^2$, $\alpha = 2\cdot10^{-3}H_0^2$, and $r_0=7\cdot 10^{-4}$ for both $p=2$ and $p=3$, while $m=50H_0$ for $p=2$ and  $m=30H_0$ for $p=3$.}
	\label{fig:p_init_arrow}
\end{figure}

For the power-law field-space metric and the potential \eqref{eq:potential}, the dark energy equation of state \eqref{eq:general_eq_of_state} becomes
\begin{align}
	\label{eq:w_eq_p_any}
	 w_{\phi}= -1 + \frac{2}{1+ \frac{p}{2} + \frac{p (V_0-\alpha\theta)}{m^2\req^2}}\,,
\end{align}
where we have assumed $\req \gg r_0$. For the specific case of $p=2$, we obtain
\begin{align}
	\label{eq:w_eq_p_2}
	 w_{\phi}= -1 + \frac{1}{1+\frac{\sqrt{3}(V_0-\alpha\theta)^{3/2}}{m\alpha }}\,,
\end{align}
where we have used the approximate expression \eqref{eq:p_metric_req} for $\req$. This analytic equation of state agrees very well with our numerical solutions as the blue dot shows in the upper left panel of \cref{fig:p_init_arrow}, where we have depicted the evolution of the fields for different $\Omega_{\phi}$ and $w_{\phi}$ initial conditions. We have allowed the fields to evolve for two $e$-folds (i.e., $\Delta N = 2$), and the arrows indicate the directions of evolution while the length of an arrow shows the relative distance that the fields have moved in the $\Omega_\phi$-$w_\phi$ phase space. The plot shows that the fields converge to the analytic semiequilibrium equation of state \eqref{eq:w_eq_p_2} marked with the blue dot. Note that the dot corresponds to $\theta=0$ in the equation and $w_\phi$ increases slowly over time as $\theta$ increases. For comparison, we have also shown in the upper right panel of \cref{fig:p_init_arrow} a phase-space evolution of the fields for the power-law metric with $p=3$. The plot shows that the approximate $w_\phi$ of \cref{eq:w_eq_p_any} for $p=3$ is slightly smaller than the numerically computed value. This can be explained by the fact that for the chosen set of parameters in the potential the kinetic terms $x_r$ and $x_\theta$ are no longer negligible and we are therefore not allowed to assume $y\approx1$, or equivalently $k_2\ll k_1$, in \cref{eq:x_theta}, which we assumed in order to obtain the analytic expression \eqref{eq:w_eq_p_any}.

The equation of state \eqref{eq:w_eq_p_any} can also be used to place analytic bounds on some of the parameters of the potential. Let us first consider the specific case of $p=2$ for which the equation of state takes the form \eqref{eq:w_eq_p_2}. We require that the denominator of the equation be large so that $w_\phi\approx -1$. We find numerically that $V_0$ is always close to the cosmological constant value of $\sim 2.19 H_0^2$ (assuming $\Omega_\phi\approx 0.7$ today), which is due to the fact that the fields are not allowed to substantially move until dark energy begins to dominate. The assumption $w_\phi\approx -1$ then results in the condition
\begin{equation}
     \label{eq:m_alpha_bound_p2}
    \frac{\alpha m}{\sqrt{3}V_0^{3/2}} \ll 1\,
\end{equation}
for $p=2$. Given that $V_0\sim 2H_0^2$ in order for the cosmological solutions to describe the observed evolution of the Universe, the condition \eqref{eq:m_alpha_bound_p2} leads to
\begin{equation}
     \label{eq:m_alpha_bound_p2_2}
    \frac{\alpha}{H_0^2}\frac{m}{H_0}\ll 5\,.
\end{equation}
More generally for any $p$, it is easy to show that \cref{eq:w_eq_p_any,eq:p_metric_req} lead to the bound
\begin{equation}
    \label{eq:m_alpha_bound_pol_general}
    2\left(\alpha^2m^p\right)^\frac{2}{p+2}V_0^{-\frac{p+4}{p+2}}\left(36p^p\right)^{-\frac{1}{p+2}}\ll 1\,,
\end{equation}
which, assuming again that $V_0\sim 2H_0^2$, leads to the simple condition
\begin{equation}
    \label{eq:m_alpha_bound_pol_general_2}
    \frac{\alpha}{H_0^2}\left(\frac{m}{H_0}\right)^{p/2} \ll 2\sqrt{3}p^{p/4}\,.
\end{equation}
\begin{figure}
	\centering
	\includegraphics[width=1\linewidth]{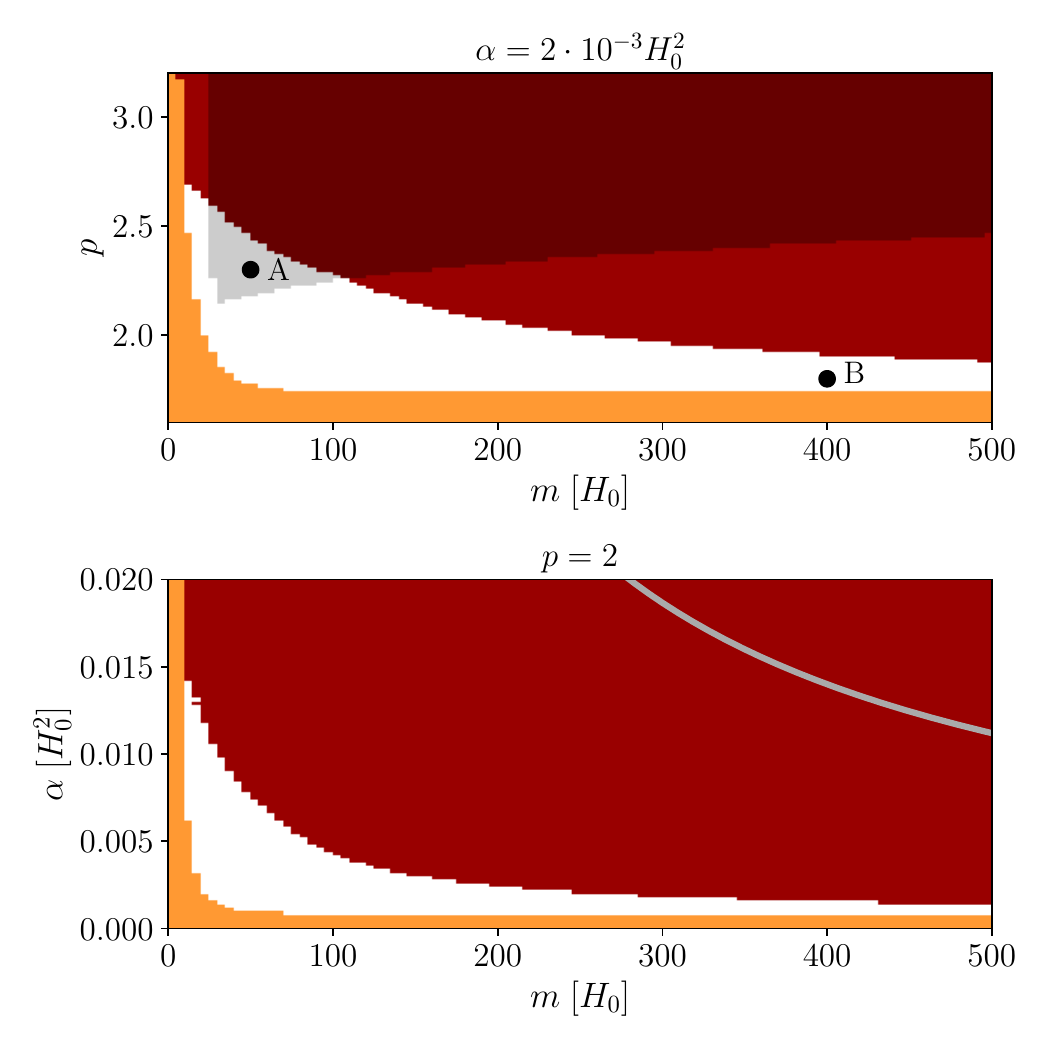}
	\caption{Qualitative exclusion plots for parameters $\alpha$, $m$, and $p$ of the potential \eqref{eq:potential} and the power-law field-space metric. The dark red regions correspond to parameter values which do not provide cosmological solutions consistent with observations of the current phase of the cosmic evolution where $\Omega_{\phi} \approx 0.7$ and $w_{\phi} \approx -1$ today, while the parameter values in the dark orange regions violate the de Sitter condition \eqref{eq:de_sitter_conjecture} with $c=0.5$ as a representative $\mathcal{O}(1)$ value. The light gray region shows the parameter values for which the speed of sound of the light mode of linear cosmological perturbations becomes imaginary at some point during the cosmic history and the perturbations are therefore plagued by gradient instabilities. We have set $\alpha=2\cdot 10^{-3}H_0^2$ in the upper panel and $p=2$ in the lower panel, while $V_0=2.19H_0^2$ and $r_0=7\cdot 10^{-4}$ for both panels. The gray solid curve in the lower panel corresponds to values of $\alpha$ and $m$ which satisfy $\alpha m =\sqrt{3}V_0^{3/2}$, showing the analytically obtained bound \eqref{eq:m_alpha_bound_p2}. The black dots A and B show the two sets of benchmark parameter values which we have used in \cref{fig:cs,fig:pert_k} for the perturbative analysis of the models with a power-law metric.}
	\label{fig:p_m_parameter}
\end{figure}Our numerical studies confirm this analytical result as shown in  \cref{fig:p_m_parameter}, where we present different numerically obtained constraints on different parameters of models with a power-law metric. The dark red regions show the excluded parts of the parameter space for which dark energy does not evolve similarly to what we observe in the Universe today, i.e., the evolution with $\Omega_\phi \approx 0.7$ and $w_\phi \approx -1$. The dark orange regions correspond to the parameter values for which the de Sitter condition \eqref{eq:de_sitter_conjecture} is violated for $c=0.5$ as a representative $\mathcal{O}(1)$ value (see \cref{sec:swamp_explicit} for details). The remaining regions therefore show the parameter values that are consistent with both cosmological observations and the de Sitter condition. We have additionally presented in the lower panel of the figure the $\alpha m =\sqrt{3}V_0^{3/2}$ curve demonstrating that our numerically obtained cosmologically viable parameter values all respect the bound \eqref{eq:m_alpha_bound_p2}, as the white region is located well below the curve.
\begin{figure}
	\centering
	\includegraphics[width=1\linewidth]{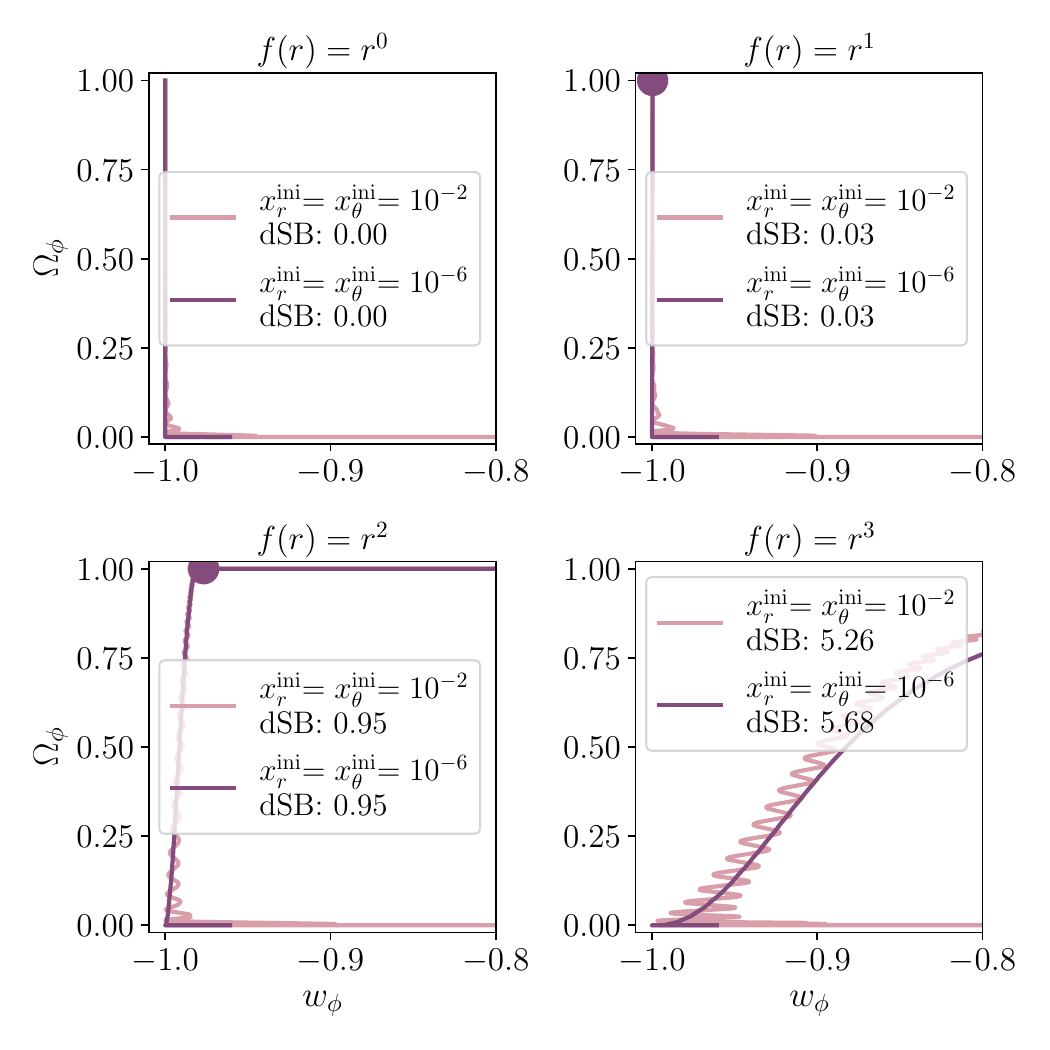}
	\caption{Examples of phase-space evolution of dark energy fields in terms of the dark energy fractional density parameter $\Omega_{\phi}$ and equation of state $w_{\phi}$ for the power-law field-space metric with $p=0,1,2,3$. Each diagram shows two trajectories corresponding to two initial values of $x_{r}$ and $x_{\theta}$. We have set $\theta^\mathrm{ini}=0$, $r^\mathrm{ini}=r_0$, and $y^\mathrm{ini}=10^{-5}$ in all cases. We have also assumed that cosmic evolution starts from a matter-dominated phase, and we have set $V_0=2.19H_0^2$, $\alpha = 10^{-3}H_0^2$, $m=50 H_0$, and $r_0 = 7\cdot 10^{-4}$. The dSB for each trajectory indicates the largest value of the constant $c$ in the swampland de Sitter condition \eqref{eq:de_sitter_conjecture} allowed by the corresponding trajectory. The colored dots correspond to the moments in the cosmic evolution by which the fields have traveled the Planckian distance $\Delta \phi = 1$. Note that for $p=0$, $r$ converges to its VEV $r_0$, making the equation of state converge to $-1$.}
	\label{fig:p_init}
\end{figure}
\begin{figure}
	\centering
	\includegraphics[width=1\linewidth]{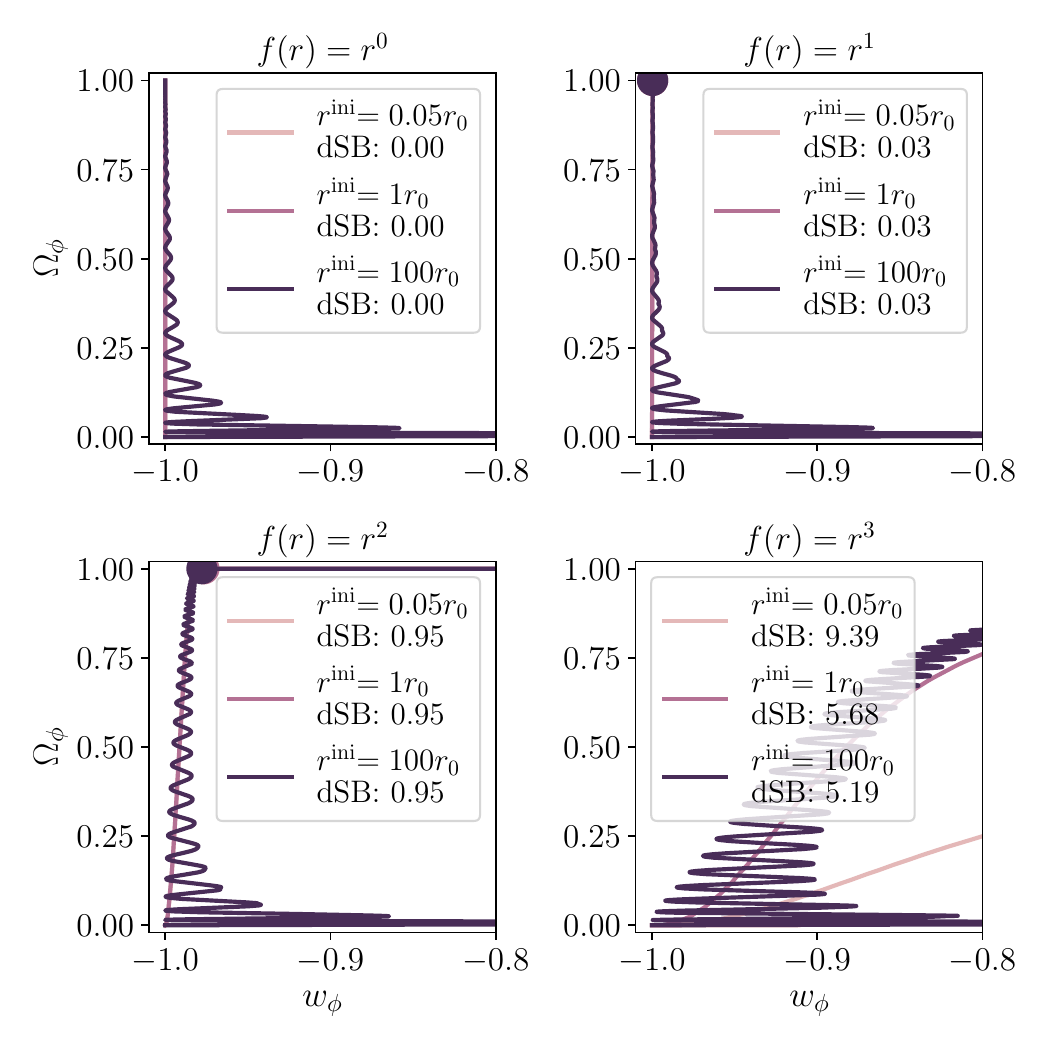}
	\caption{As in \cref{fig:p_init}, but for three different initial values of $r$. In all cases we have set $\theta^\mathrm{ini}=0$, $x^\mathrm{ini}_r =0$, $x^\mathrm{ini}_{\theta}=0$ and $y^\mathrm{ini}=10^{-5}$.}
	\label{fig:p_r_init}
\end{figure}

It is important to note that \cref{fig:p_m_parameter} is only a qualitative representation of how the parameter space of a model can be constrained by observational and theoretical considerations, and precise constraints can only be provided through a rigorous statistical exploration of the parameter space. Figure~\ref{fig:p_m_parameter}, however, provides a useful illustration of how the dark energy models we study in this paper can be strongly constrained when a combination of theoretical and observational constraints is considered.

It is also interesting to know how the evolution of scalar fields depends on different initial conditions. In \cref{fig:p_init}, we again show phase-space diagrams for the power-law field-space metric in terms of $\Omega_\phi$ and $w_\phi$, as we did in \cref{fig:p_init_arrow}. Here, however, we only show, for each diagram, two cosmic trajectories corresponding to two different sets of initial conditions for the field velocities $x_r$ and $x_{\theta}$: $x^\mathrm{ini}_r=x^\mathrm{ini}_{\theta}=10^{-2}$ and $x^\mathrm{ini}_r=x^\mathrm{ini}_{\theta}=10^{-6}$. We present the diagrams for four cases of $r^0$, $r^1$, $r^2$, and $r^3$, and we set $\theta^\mathrm{ini}=0$, $r^\mathrm{ini}=r_0$, and $y^\mathrm{ini}=10^{-5}$ for the initial values of $\theta$, $r$, and $y$. The figure shows oscillations along the trajectories for the initial conditions $x^\mathrm{ini}_r=x^\mathrm{ini}_{\theta}=10^{-2}$. This is because the $r$ field oscillates while climbing up the potential, as seen in \cref{fig:polar} for the representative case of $r^2$. There are similar oscillations for the other case of $x^\mathrm{ini}_r=x^\mathrm{ini}_{\theta}=10^{-5}$, but the oscillations are too small to be seen in the figure. Our numerical analysis shows similar trajectories even if one of the two quantities $x^\mathrm{ini}_r$ and $x^\mathrm{ini}_{\theta}$ is set to zero. We also notice that each trajectory shown in \cref{fig:p_init} converges to the analytic value of the equation of state give by \cref{eq:w_eq_p_any} for the corresponding value of $p$. These asymptotic equations of state increase with time very slowly as $\theta$ increases and falls down the potential. All these observations suggest that for the potential \eqref{eq:potential} and the power-law metric $r^p$ the overall evolution of the fields is not highly sensitive to the initial conditions for $\dot{r}$ and $\dot{\theta}$.

Similarly, \cref{fig:p_r_init} shows that the evolution of the system does not depend strongly on the initial value of $r$. The figure presents the phase-space dynamics for the same field-space metrics and values of the parameters $V_0$, $\alpha$, $m$, and $r_0$ as in \cref{fig:p_init}. For each of the four diagrams corresponding to the four metrics, we show three cosmic trajectories for three initial values of $r$: $0.05r_0$, $r_0$, and $100r_0$. In all these cases, we have set $\theta^\mathrm{ini}=0$, $x_r^\mathrm{ini}=0$, $x_\theta^\mathrm{ini}=0$, and $y^\mathrm{ini}=10^{-5}$ for the initial values of $\theta$, $x_r$, $x_\theta$, and $y$. Similar to \cref{fig:p_init}, here we see that even though the exact trajectories differ for different initial values of $r$, all of them converge asymptotically to $\Omega_\phi=1$ and the analytic value of $w_\phi$ given by \cref{eq:w_eq_p_any}. One case which may seem to be behaving differently is the $r^3$ case with the small initial value for $r$, i.e., the $r^\mathrm{ini}=0.05r_0$ case in the lower right panel of \cref{fig:p_r_init}. Our numerical analysis shows, however, that even though $\Omega_\phi$ increases much more slowly in that case compared to the cases with $r^\mathrm{ini}\geq r_0$, the trajectory eventually converges to the same point in the phase space as for the other trajectories.

We do not need to discuss in detail how cosmological solutions depend on the initial value of $\theta$, as changing $\theta^\mathrm{ini}$ is simply equivalent to changing the value of the parameter $V_0$ in the potential.

\subsection{Hyperbolic field-space metric}
\label{sec:hyp-metric}
We now consider a field-space metric with $f(r)=e^{\beta r}$ and the same potential as in the previous section, i.e., the potential \eqref{eq:potential}, and we set $r_0=0$ without loss of generality. Both the metric and the potential are now $\mathbb{Z}_2$-symmetric, i.e., the transformation 
\begin{equation}
    \begin{cases}
    \beta \rightarrow -\beta\\
    r \rightarrow -r
    \end{cases}
\end{equation}
leaves the action invariant. This means that we are allowed to consider both positive and negative values of $\beta$. In the rest of this paper, we assume that $\beta$ is positive, but because of the $\mathbb{Z}_2$ symmetry, all of our results are also valid for negative $\beta$.

\begin{figure}
	\centering
	\includegraphics[width=1\linewidth]{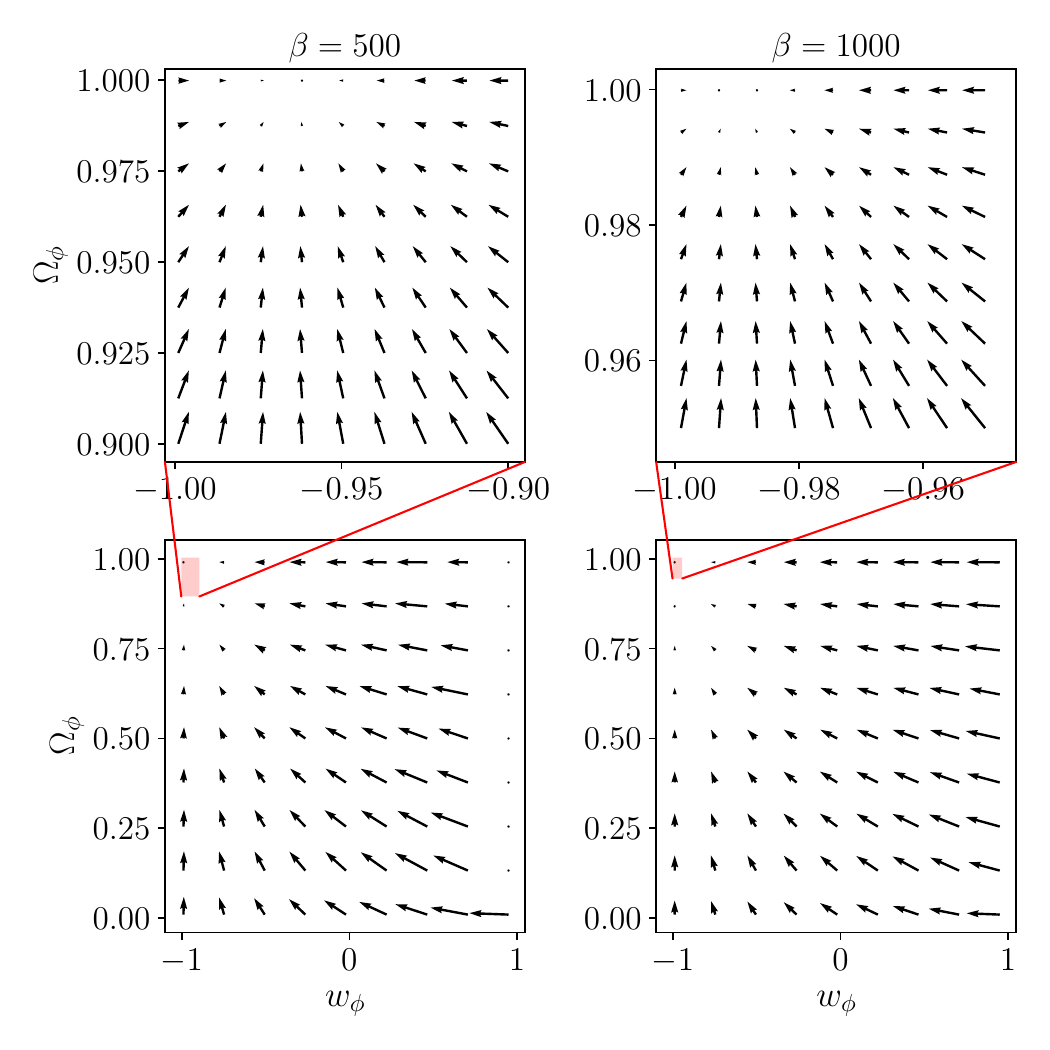}
	\caption{As in \cref{fig:p_init_arrow}, but for the hyperbolic field-space metric with $\beta=500$ and $\beta=1000$. We have set $V_0=2.19H_0^2$, $\alpha = 3H_0^2$, $m=50H_0$ and $r_0=0$ in both cases.}
	\label{fig:exp_init_arrow}
\end{figure}

For the hyperbolic metric, \cref{eq:turningrate_general} for the turning rate gives
\begin{equation}
	\label{eq:turningrateexp}
	\Omega^2 = \frac{1}{2} m^2 \beta \req\,,
\end{equation}
where we have assumed that the $r$ field is at its semiequilibrium value $\req$, which satisfies the equation
\begin{equation}
\label{eq:req_equation_hyper}
\req^3 + \frac{2}{m^2}\left(V_0 - \alpha \theta \right)\req =  \frac{\alpha^2}{3m^4}\beta e^{-\beta \req}\,.
\end{equation}
This equation does not have a closed-form solution for $\req$, but we immediately see that $\req$ is positive for positive $\beta$. This is assuming that $V_0-\alpha\theta$ is positive for the entire history of the Universe, including late times where dark energy becomes dominant. The approximate equation of state \eqref{eq:general_eq_of_state} now becomes
\begin{align}
	w_{\phi} &= -1 + \frac{2}{1+\beta \frac{V}{V_r}}\,\nonumber\\
	&= -1+\frac{2}{1+\frac{\beta}{2}\req+\frac{\beta(V_0-\alpha\theta)}{m^2\req}}\,.\label{eq:w_hyper}
\end{align}
Since here, contrary to the power-law field-space metric of the previous section, we do not have an analytical solution for $\req$, we cannot provide an approximate expression for the asymptotic dark energy equation of state similar to \cref{eq:w_eq_p_2} for the $p=2$ power-law metric. However, the $\Omega_\phi$-$w_\phi$ phase-space diagrams of \cref{fig:exp_init_arrow} show that the hyperbolic metric also leads to semiattractor solutions.\footnote{See \cref{sec:swamp_explicit} for an explanation of why in the figure we have set $\alpha$ to an $\mathcal{O}(H_0^2)$ value.} For the examples of $\beta=500$ and $\beta=1000$ in the figure, the asymptotic points are positioned at $\{w_\phi \approx -0.96, \Omega_\phi\approx 1\}$ and $\{w_\phi \approx -0.99, \Omega_\phi\approx 1\}$, respectively.

\begin{figure}
	\centering
	\includegraphics[width=1\linewidth]{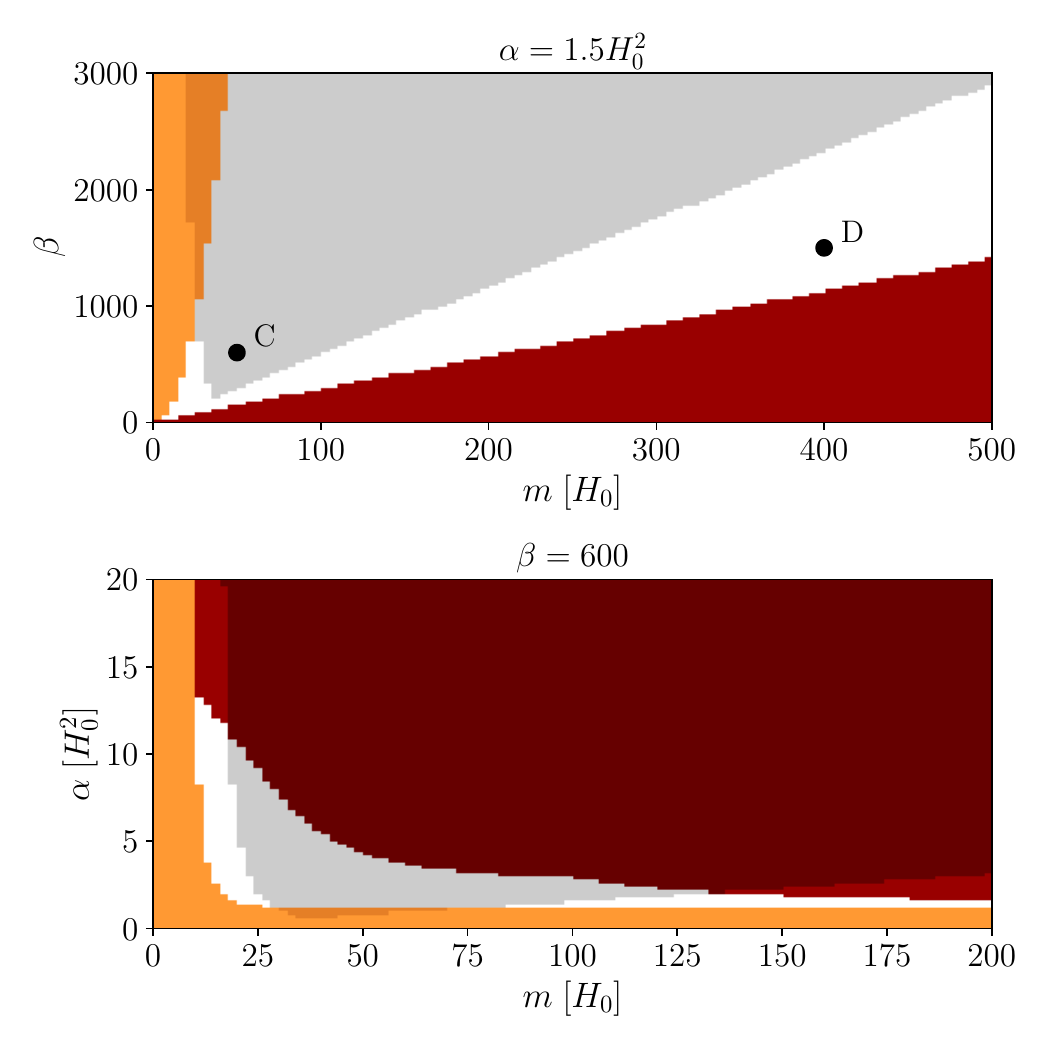}
	\caption{As in \cref{fig:p_m_parameter}, but for the hyperbolic field-space metric. We have set $\alpha=1.5H_0^2$ in the upper panel and $\beta=600$ in the lower panel. In both panels we have set $V_0=2.19H_0^2$ and $r_0=0$.}
	\label{fig:m_beta_param}
\end{figure}

By requiring the equation of state \eqref{eq:w_hyper} to be close to $-1$, we obtain the condition
\begin{equation}
	\frac{\beta (V_0-\alpha\theta)}{m^2 \req} + \frac{1}{2}\beta \req \gg 1\,.\label{eq:cond_hyper}
\end{equation}
On the other hand, \cref{eq:req_equation_hyper} implies that $\beta\req$ cannot be much larger than unity. We can see this by multiplying the equation by $\beta^3$,
\begin{equation}
     \label{equilibrium_exp}
	(\beta \req)^3 + \frac{2\beta^2}{m^2}\left(V_0 - \alpha \theta \right)(\beta\req) =  \frac{\alpha^2\beta^4}{3m^4} e^{-\beta \req}\,.
\end{equation}
Now if we assume $\beta \req \gg 1$, and given that $V_0-\alpha\theta$ is positive as mentioned earlier, the right-hand side of \cref{equilibrium_exp} quickly tends to zero and the only real solution for $\beta\req$ will be $\beta \req \rightarrow 0$, which violates our assumption of $\beta \req \gg 1$. Since we cannot have $\beta \req \gg 1$, the condition \eqref{eq:cond_hyper} then leads to the requirement
\begin{equation}
    \beta \gg \left(\frac{m}{H_0} \right)^2 \req\,,
\end{equation}
as $V_0$ is of $\mathcal{O}(H_0^2)$ and $\alpha\theta\ll V_0$. By combining this condition with the requirement that the turning rate \eqref{eq:turningrateexp} must satisfy $\Omega \gg H$ in order for steep potentials to provide viable cosmic acceleration, we obtain the condition
\begin{equation}
    \label{eq:exp_beta_inequality}
	\beta^2\gg \left(\frac{m}{H_0} \right)^2 \beta\req \gg \left(\frac{H}{H_0} \right)^2\,.
\end{equation}
Assuming that today (i.e., when $H=H_0$) the field $r$ is almost at its semiequilibrium value $\req$, the condition \eqref{eq:exp_beta_inequality} leads particularly to
\begin{equation}
    \label{eq:exp_beta_inequality_2}
	\left(\frac{m}{H_0} \right)^2 \beta\req \gg 1\,.
\end{equation}
Since $\beta \req$ cannot be much larger than unity, this condition provides a lower bound on the mass of the $r$ field,
\begin{equation}
    \label{eq:exp_beta_inequality_3}
	m\gg H_0\,,
\end{equation}
while the condition \eqref{eq:exp_beta_inequality} also implies that $\beta\gg1$. Note that the condition \eqref{eq:exp_beta_inequality_3} comes from the assumption that $\Omega$ is large, and therefore, in addition to being necessary for steep potentials to work, it is also a necessary condition for the models to satisfy the de Sitter swampland condition \eqref{eq:de_sitter_conjecture}. Our numerical results confirm this as shown in \cref{fig:m_beta_param}, where numerically obtained constraints on different parameters of models with a hyperbolic metric are presented for two cases of fixed $\alpha$ and $\beta$; see \cref{fig:p_m_parameter} for the description of different exclusion regions. The dark orange regions, which correspond to the parts of the parameter space which do not satisfy the de Sitter condition, basically show the parameter values which do not provide large turning rates. It is then clear from the figure that small values of $m$ are excluded by that constraint when $\beta\gg 1$, which is in agreement with the condition \eqref{eq:exp_beta_inequality}. The dark red region in the upper panel of \cref{fig:m_beta_param} also shows that for any given value of $m$ (and $\alpha$) there is a lower bound on the parameter $\beta$. Our numerical analysis shows that a large value of $\beta$ forces $\req$ to become small and scale approximately as $\req \propto 1/\beta$. We show in an accompanying paper \cite{Eskilt2022} that a large $\beta$ has strong effects on the evolution of cosmological perturbations.

\begin{figure}
	\centering
	\includegraphics[width=1\linewidth]{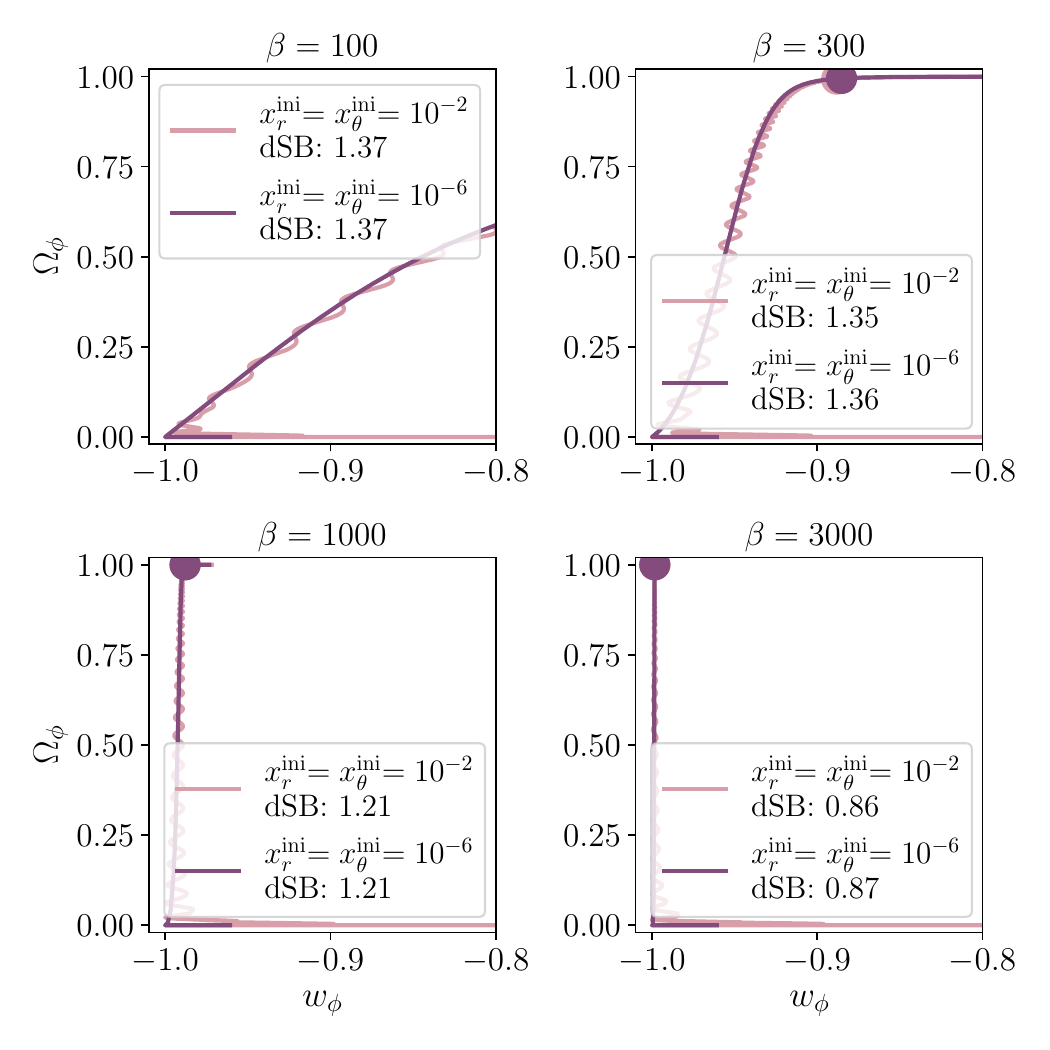}
	\caption{As in \cref{fig:p_init}, but for the hyperbolic field-space metric with $\beta = 100$, $\beta=300$, $\beta=1000$, and $\beta=3000$. In all cases we have set $\theta^\mathrm{ini}=0$, $r^\mathrm{ini}=r_0$, and $y^\mathrm{ini}=10^{-5}$. We have also set $V_0=2.19H_0^2$, $\alpha = 3 H_0^2$, $m=50 H_0$, and $r_0 = 0$.}
	\label{fig:exp_init}
\end{figure}

\begin{figure}
	\centering
	\includegraphics[width=1\linewidth]{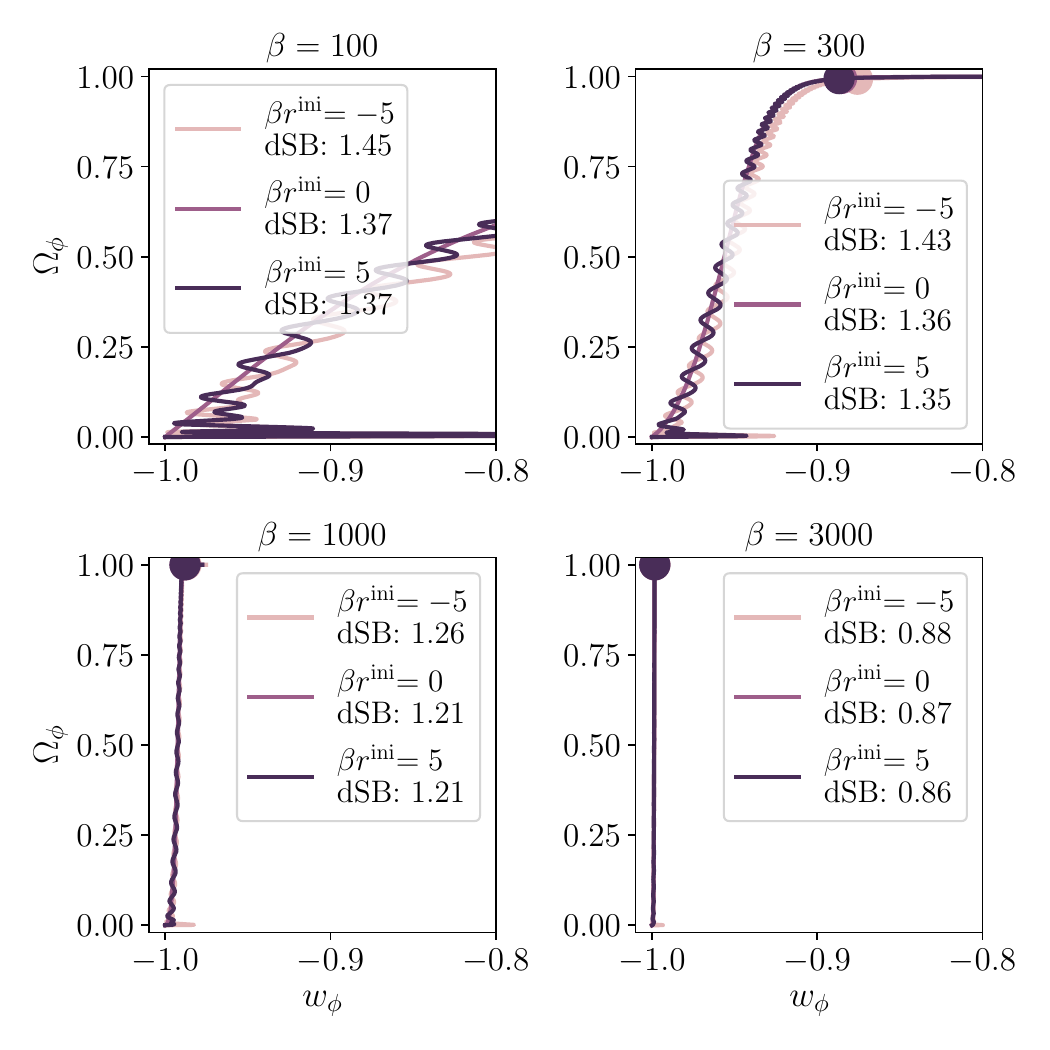}
	\caption{
	As in \cref{fig:exp_init}, but for three different initial values of $r$. In all cases we have set $\theta^\mathrm{ini}=0$, $x^\mathrm{ini}_r =0$, $x^\mathrm{ini}_{\theta}=0$, and $y^\mathrm{ini}=10^{-5}$.}
	\label{fig:exp_r_init}
\end{figure}

As in the case of the power-law field-space metric, it is interesting to know how the evolution of the scalar fields $r$ and $\theta$ depends on the initial values of $r$ and the field velocities $x_r$ and $x_\theta$. In \cref{fig:exp_init}, we present four diagrams, each showing two $\Omega_\phi$-$w_\phi$ phase-space trajectories corresponding to the two sets of initial conditions $x^\mathrm{ini}_r=x^\mathrm{ini}_{\theta}=10^{-2}$ and $x^\mathrm{ini}_r=x^\mathrm{ini}_{\theta}=10^{-6}$, as we did in \cref{fig:p_init} for the power-law metric. The diagrams correspond to four cases of $\beta = 100$, $\beta=300$, $\beta=1000$ and $\beta=3000$, where we have set $\theta^\mathrm{ini}=0$, $r^\mathrm{ini}=0$, and $y^\mathrm{ini}=10^{-5}$ for the initial values of $\theta$, $r$, and $y$. Similar to the power-law metric of the previous section, the larger the initial values of $x_r$ and $x_\theta$, the larger the oscillations along the trajectories caused by oscillations in the field $r$. Finally, we present in \cref{fig:exp_r_init} four similar diagrams, each showing three phase-space trajectories corresponding to the three sets of initial conditions $r^\mathrm{ini}=-5/\beta$, $r^\mathrm{ini}=0$ and $r^\mathrm{ini}=5/\beta$ for the field $r$. In all these cases, we have set $\theta^\mathrm{ini}=0$, $x_r^\mathrm{ini}=0$, $x_\theta^\mathrm{ini}=0$, and $y^\mathrm{ini}=10^{-5}$ for the initial values of $\theta$, $x_r$, $x_\theta$, and $y$, as we did in \cref{fig:p_r_init} for the power-law metric. Comparing  \cref{fig:exp_init,fig:exp_r_init} with \cref{fig:p_init,fig:p_r_init} demonstrates that the hyperbolic field-space metric shares many features with the power-law metric in terms of the background evolution of dark energy fields.

\subsection{Swampland constraints}
\label{sec:swamp_explicit}

In this section, we study the swampland conjectures of \cref{sec:swampland} in more detail and in context of the explicit models of power-law and hyperbolic field-space metrics with the potential \cref{eq:potential}. We are particularly interested in the theoretical constraints that the conjectures would place on the parameters of the models.

Let us assume, without loss of generality, that the $\theta$ field is initially set to zero.\footnote{Any other initial values of $\theta$ are equivalent to shifts in the parameter $V_0$ of the potential given that the field-space metric and therefore the kinetic terms in the action \eqref{eq:action} are independent of $\theta$.} Our analysis shows that independent of the initial value of the $r$ field, it quickly and almost instantly drops into its vacuum expectation value $r_0$ before $\theta$ moves substantially. The field $r$ then moves up the potential, away from its VEV and toward the semiequilibrium value $\req$ as $\theta$ increases; see, e.g., \cref{fig:polar} for the examples with the power-law metric $r^2$. This means that the field $r$ takes the value $r_0$ at some point during the early evolution of the fields and when $\theta\approx 0$.

It is then easy to show that for the field $r$ sitting at its vacuum expectation value $r_0$, the de Sitter condition \eqref{eq:de_sitter_conjecture} becomes 
\begin{equation}\label{eq:dScond_quadV}
   \frac{\alpha }{\sqrt{f(r_0)}V_0} \geq c\,
\end{equation}
for any field-space metric of the form \eqref{eq:metric} with a positive $f(r,\theta)=f(r)$, and assuming $V_0>0$ and $\theta\approx 0$. Restricting ourselves to the classes of metrics studied in the previous sections, the condition then becomes
\begin{equation}
     r_0^{-\frac{p}{2}}\frac{\alpha}{V_0}\geq c\,\label{eq:swampland_dS_pol}
\end{equation}
for the power-law metric, where $f(r) = r^p$, and
\begin{equation}
  \frac{\alpha }{V_0}\geq c\,\label{eq:swampland_dS_hyper}
\end{equation}
for the hyperbolic metric, where $f(r) = e^{\beta r}$. Note that we have set $r_0=0$ for the hyperbolic metric, while we assume $0 < r_0 \ll 1$ for the power-law metric, as it does not allow $r_0=0$. For a $V_0$ of $\mathcal{O}(H_0^2)$, i.e., the order of the dark energy scale, and a $c$ of $\mathcal{O}(1)$, the condition  \eqref{eq:swampland_dS_hyper} implies that $\alpha$ cannot be smaller than $\mathcal{O}(H_0^2)$ for the hyperbolic metric, while the condition \eqref{eq:swampland_dS_pol} implies that it can be arbitrarily small for the power-law metric, as $r_0$ can be as close to zero as we want.

\begin{figure}
	\centering
	\includegraphics[width=1\linewidth]{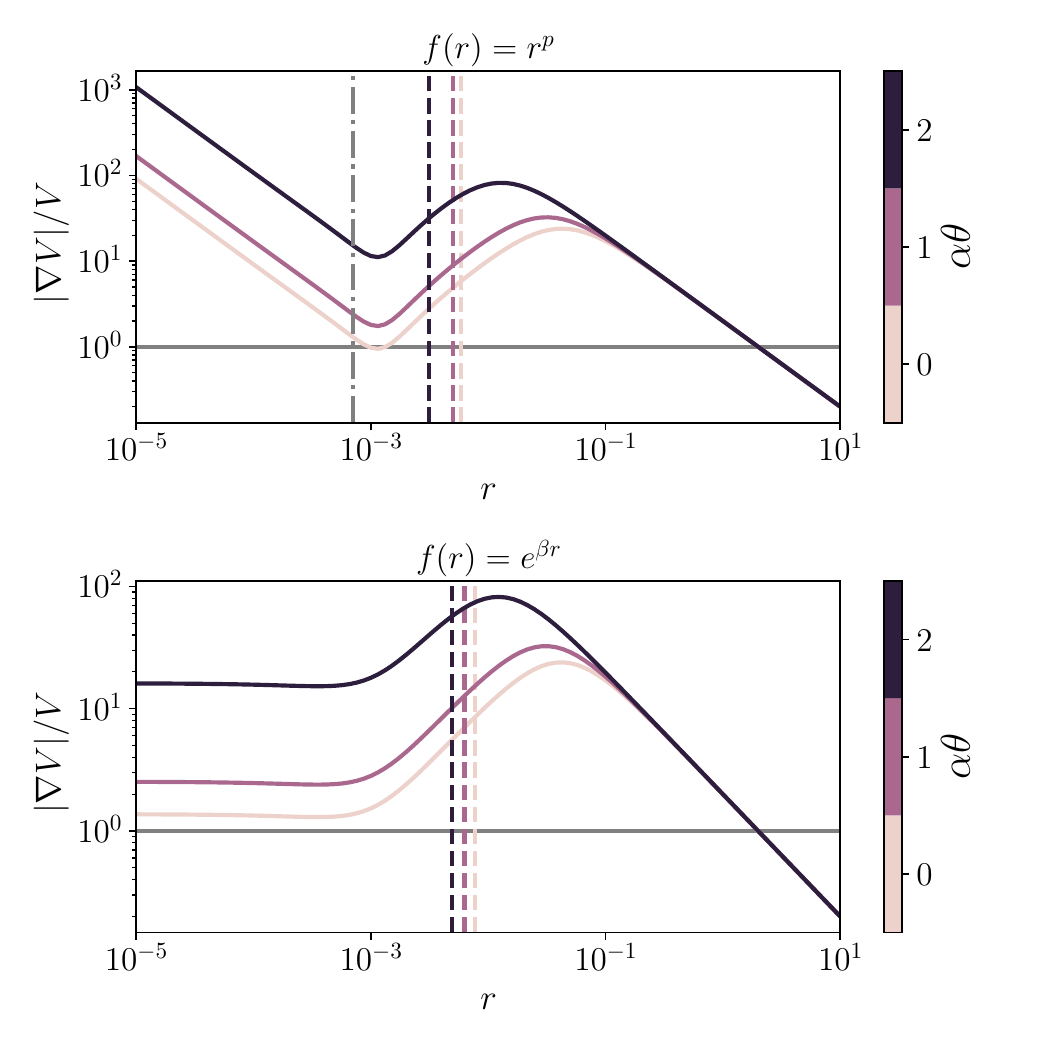}
	\caption{The de Sitter conjecture quantity $|\nabla V|/V$ as a function of the fields $r$ and $\alpha\theta$ for the power-law (upper panel) and hyperbolic (lower panel) field-space metrics, with $p=2$ and $\beta=600$, respectively. The dashed vertical lines show the semiequilibrium value $\req$ for the $r$ field, and the gray dashed-dotted vertical line shows $r_0=7 \cdot 10^{-4}$ for the power-law metric. We have set $r_0=0$ for the hyperbolic metric. $V_0=2.19 H_0^2$ and $m=50H_0$ in both panels. The horizontal line corresponds to $|\nabla V|/V=1$.}
	\label{fig:de_sitter}
\end{figure}

Figure \ref{fig:de_sitter} shows examples of how the de Sitter conjecture quantity $|\nabla V|/V$, or the slope of the potential, varies with respect to $r$ and $\theta$. Note that we show the curves as a function of $\alpha\theta$ instead of $\theta$, as it is that combination which appears in $|\nabla V|/V$. We have set $V_0=2.19 H_0^2$ and $m=50H_0$ in both cases, while $\alpha = 2\cdot 10^{-3} H_0^2$ and $r_0 = 7\cdot 10^{-4}$ for the power-law metric, and $\alpha=3 H_0^2$ and $r_0=0$ for the hyperbolic metric; these are the values we used for the examples of the previous sections. The figure shows that for these values of the parameters, the de Sitter condition is not violated as long as $r \ll 10$, which is guaranteed if $r$ starts with a small value initially. The reason is that independent of the initial value of $r$, it almost instantly goes to its VEV, $r_0$, and then moves asymptotically toward its semiequilibrium value $\req$, which is much smaller than $10$ in the examples of \cref{fig:de_sitter}. As briefly mentioned in \cref{sec:pol,sec:hyp-metric}, we have provided in the exclusion plots of \cref{fig:p_m_parameter,fig:m_beta_param} examples of the regions in the parameter space for both power-law and hyperbolic metrics which violate the de Sitter condition \eqref{eq:de_sitter_conjecture}. These are the dark orange regions in the figures for which we have assumed $c=0.5$ as a representative example of $\mathcal{O}(1)$ values. We discussed some of the features of these excluded regions in \cref{sec:pol,sec:hyp-metric} in terms of the analytical conditions imposed on the parameters of the models by assuming that the turning rate $\Omega$ is much larger than the Hubble expansion rate. Note that, as we discussed in those sections, not only is this assumption of $\Omega\gg H$ necessary for steep potentials to provide desired cosmological solutions, it is also effectively equivalent to requiring that the de Sitter swampland condition is satisfied. 

In \cref{fig:p_init,fig:p_r_init,fig:exp_init,fig:exp_r_init} of the previous sections, where we presented examples of the phase-space trajectory of dark energy fields for different power-law and hyperbolic metrics and for different initial conditions, we have also provided the largest value of the constant $c$ that is allowed by each trajectory. These are the values labeled by ``dSB'' (for de Sitter Bound) in the figures and are obtained by computing the quantity
\begin{equation}
\textrm{min}\left(\frac{|\nabla V|}{V} \right)
\end{equation}
over the entire field evolution of each trajectory. 
The figures show that all the trajectories satisfy the de Sitter swampland conjecture with a $c=\mathcal{O}(1)$, except the ones corresponding to power-law metrics with $p=0$ and $p=1$, for which \cref{fig:p_init,fig:p_r_init} show that the de Sitter condition is violated at some point during the evolution of the fields.

In the same \cref{fig:p_init,fig:p_r_init,fig:exp_init,fig:exp_r_init}, we have also provided for each trajectory in the $\Omega_\phi$-$w_\phi$ phase space the moment in the cosmic evolution by which the fields have traveled the Planckian distance $\Delta \phi = 1$, where $\Delta \phi$ is defined by \cref{eq:distance1}. These moments, which are marked by colored dots in the figures, are those beyond which the effective field theory breaks down according to the swampland distance conjecture; see \cref{sec:swampland}.
The figures show that for the models studied in the previous sections, the breakdown of the effective field theory does not happen in the near future as $\Omega_\phi\sim 1$ at those points, corresponding to the far future.

\section{Sound speed of linear perturbations and gradient instability}

In this section, we investigate constraints that a simple theoretical analysis of linear cosmological perturbations may additionally place on our models of multifield dark energy and their free parameters. Here, we do not intend to perform a detailed and extensive analysis of the perturbations and restrict ourselves to a regime of linear scales where certain theoretical approximations significantly simplify the analysis. We provide a rigorous and exhaustive study of cosmological perturbations in an accompanying paper \cite{Eskilt2022} where we explore various theoretical features and observational implications of the models for large-scale structure surveys.

As shown in \rcite{Akrami:2020zfz}, in the subhorizon regime, i.e., on comoving scales $k$ where $H^2a^2\ll k^2$, the cosmological linear perturbations of the two-field dark energy models described by the action \eqref{eq:action} contain a light mode which propagates with a sound speed $c_\mathrm{s}$ given by
\begin{equation}
    \label{eq:cs2_general}
    c_\mathrm{s}^{2} = \frac{1}{1+ \frac{4a^2\Omega^2}{M_\mathrm{eff}^2}}\,,
\end{equation}
where $M_\mathrm{eff}$ is an effective mass given by
\begin{equation}
	M_\mathrm{eff}^2 = a^2 \left(V_{\mathcal{N}\mathcal{N}} - \Omega^2 +\mathcal{R}\frac{{\dot{\phi}}^2}{2}\right)\,.\label{eq:Meff2-general}
\end{equation}
Here, $V_\mathcal{NN} \equiv \mathcal{N}^a\mathcal{N}^b\mathcal{D}_a\mathcal{D}_b V$, where $\mathcal{N}$ is the normalized normal vector to the field-space trajectory given by
\begin{equation}\label{eq:normal}
    \mathcal{N}^a=-\frac{1}{\Omega}D_t\mathcal{T}^a=-\frac{1}{\Omega}\frac{\dot\phi^a}{\dot\phi}\,
\end{equation}
and $\mathcal{D}_a$ is the covariant derivative associated with the field-space metric $\mathcal{G}_{ab}$. $\mathcal{R}$ is the Ricci scalar for $\mathcal{G}_{ab}$.

The linear perturbations also include a heavy mode, and the expression \eqref{eq:cs2_general} for the sound speed of the light mode is valid only on scales larger than the Compton wavelength of the heavy mode, $k^2\ll M_\mathrm{eff}^2c_\mathrm{s}^{-2}$, where the heavy mode can be integrated out, leading to a well-defined single-field effective theory. Combining this theoretical constraint on the scales with the assumption that they are subhorizon, we can then use \cref{eq:cs2_general} to analytically investigate the clustering properties of dark energy and the presence or absence of gradient instabilities for perturbations over the range
\begin{equation}
    H^2a^2\ll k^2\ll M_\mathrm{eff}^2c_\mathrm{s}^{-2}\,.\label{eq:k_range}
\end{equation}

As argued in \rcite{Akrami:2020zfz}, since the turning rate $\Omega$ can be very large in our multifield models, \cref{eq:cs2_general} implies that the sound speed of the light mode can consequently be suppressed when $a^2\Omega^2\gg M_\mathrm{eff}^2$, and therefore, be considerably smaller than unity, causing dark energy to cluster on scales larger than the dark energy sound horizon.\footnote{Note that this sound horizon is not the same as the cosmological horizon and can be much smaller than that, leading to dark energy clustering at observable subhorizon scales.}

Let us now focus on the explicit examples of two-field dark energy models studied in the previous sections, where the potential is of the form~\eqref{eq:potential} and the field-space metric is either power-law or hyperbolic, with $f(r)=r^p$ or $f(r)=e^{\beta r}$, respectively. As discussed before, for appropriate choices of free parameters, the $r$ field of the models reaches a semiequilibrium value $\req$ at late times during the cosmic evolution after which it varies slowly with time, i.e., $\dot{r} \sim 0$. It is easy to show that in this limit
\begin{align}
    V_\mathcal{NN} &\approx V_{rr}(\req)\,,\\
    \dot{\phi}^2 &\approx f(\req)\dot{\theta}^2 \approx 2\frac{ f(\req)}{f_r(\req)}V_r(\req)\,,
\end{align}
where we have used the definition \eqref{eq:phidotsqr} for $\dot{\phi}^2$ and the $r$-field equation of motion \eqref{eq:r_eq_of_mot}. Using these approximations for $V_\mathcal{NN}$ and $\dot{\phi}^2$, as well as \cref{eq:turningrate_general} for $\Omega^2$, we obtain an approximate analytic form for $\Meff^2$,
\begin{equation}
    \Meff^2 \approx a^2 \left(V_{rr} - \frac{1}{2}\frac{f_r}{f}V_r + \mathcal{R} \frac{f}{f_r}V_r \right)\bigg|_{r=\req}\,,\label{eq:Meff2-approx}
\end{equation}
where the field-space Ricci scalar $\mathcal{R}$ is
\begin{equation}
    \mathcal{R}=-\frac{1}{f}\left(f_{rr}-\frac{1}{2}\frac{f_r^2}{f}\right)\,.
\end{equation}

\begin{figure}
	\centering
	\includegraphics[width=1\linewidth]{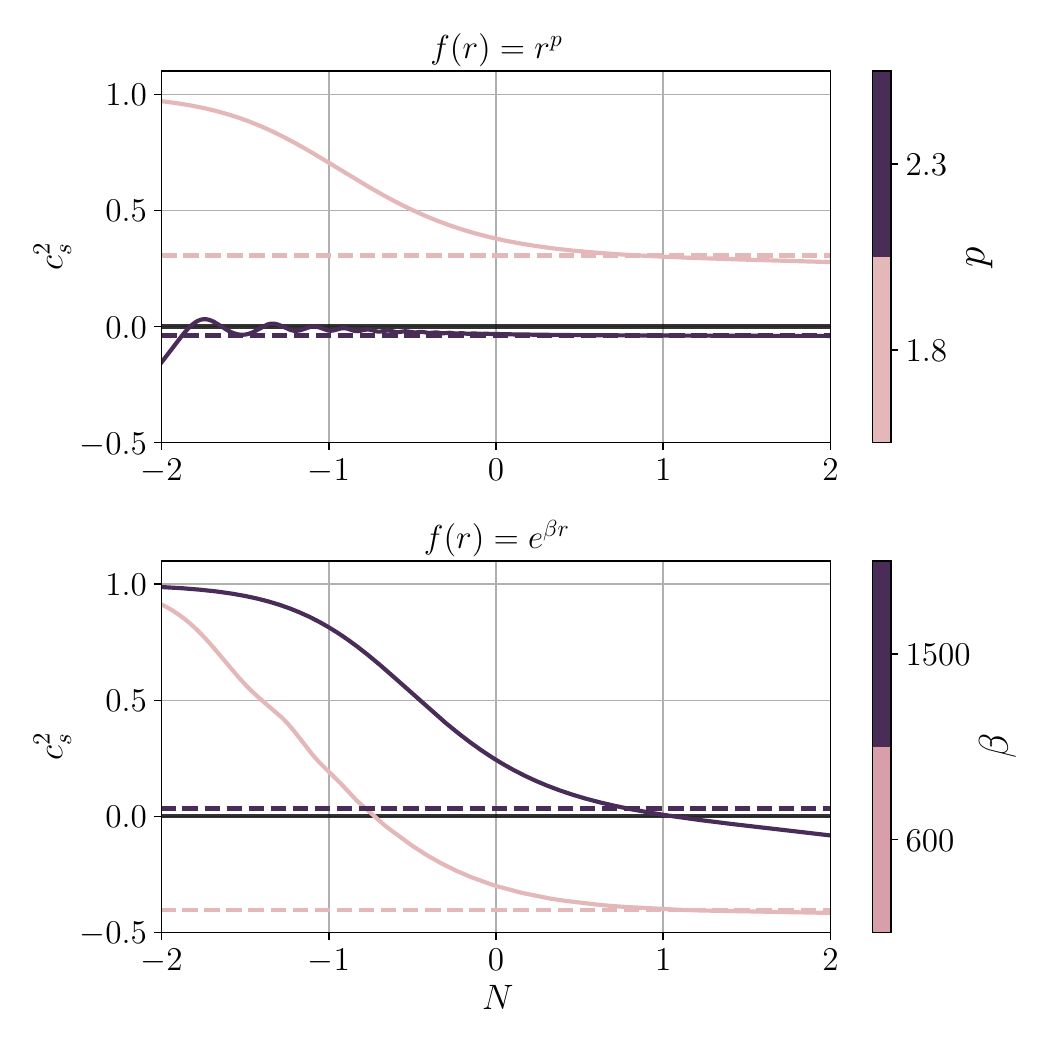}
	\caption{Time evolution of the sound speed of cosmological linear perturbations (for the light propagating mode) in terms of the number of $e$-folds $N$ for power-law (upper panel) and hyperbolic (lower panel) field-space metrics with different values of $p$ and $\beta$ ($N=0$ corresponds to the present time). The solid curves show the exact, numerically computed values of $c_\mathrm{s}^2$, while the corresponding dashed horizontal lines show the analytic approximations to the quantities computed at the semiequilibrium values $\req$ of the $r$ field, i.e., \cref{eq:c_s_s_p} for the upper panel and \cref{eq:cs2_exp} for the lower panel. Upper panel: $m=400H_0$ and $m=50H_0$ for $p=1.8$ and $p=2.3$, respectively, while $\alpha =2\cdot 10^{-3} H_0^2$ and $r_0 = 7\cdot 10^{-4}$ for both cases.  Lower panel: $m=50H_0$ and $m=400H_0$ for $\beta=600$ and $\beta=1000$, respectively, while $V_0=2.19H_0^2$, $\alpha = 1.5 H_0^2$, and $r_0=0$ for both cases. These parameter sets are all marked with letters A, B, C, and D in \cref{fig:p_m_parameter,fig:m_beta_param}.}
	\label{fig:cs}
\end{figure}

Given that
\begin{equation}
    \mathcal{R}=-\frac{p(p-2)}{2r^2}\,
\end{equation}
for the power-law field-space metric, \cref{eq:Meff2-approx} reduces to the simple expression
\begin{equation}
    \Meff^2 \approx a^2m^2\left(2-p + \frac{r_0}{\req}\left(p-1\right) \right)\,,
\end{equation}
which results in the approximate sound speed
\begin{equation}
    \label{eq:c_s_s_p}
    c_\mathrm{s}^2 \approx \frac{2-p + (p-1)\frac{r_0}{\req}}{2+p - (p+1)\frac{r_0}{\req}}\,,
\end{equation}
where we have again used \cref{eq:turningrate_general} for $\Omega^2$. As we discussed in \cref{sec:pol}, for a wide range of parameter values with desired cosmic evolution, $\req \gg r_0$, which then leads to a further simplification of the sound speed for power-law metrics,
\begin{equation}
    \label{eq:c_s_s_p_2}
    c_\mathrm{s}^2 \approx \frac{2-p}{2+p}\,.
\end{equation}
As expected, for $p=0$, i.e., the trivially flat field-space metric, $c_\mathrm{s}\approx1$, while for the $p=2$ case, i.e., the second (and only other) flat metric, the sound speed is close to zero, showing that dark energy clusters on subhorizon scales. We also see that for $p>2$ (negative curvature) the quantity $c_\mathrm{s}^2$ is negative at late times and remains negative in the future, signaling a gradient instability. The sound speed squared is always positive for $p<2$ (positive curvature). Note, again, that all these statements are valid only for parameter values for which $\req \gg r_0$. In the upper panel of \cref{fig:cs}, we show the exact, numerically obtained $c_\mathrm{s}^2$ for two representative cases of $p=1.8$ (for $p<2$) and $p=2.3$ (for $p>2$) as a function of the number of $e$-folds $N$, with $N=0$ corresponding to the present time, as well as the approximate analytic values given by \cref{eq:c_s_s_p} in combination with the value of $\req$ obtained from \cref{eq:general_req}. The figure shows that the exact values of $c_\mathrm{s}^2$ converge to the analytic approximate values. This is expected as the analytic values are valid only when the field $r$ has reached its semiequilibrium value $\req$. The figure shows, however, that after reaching this value, $c_\mathrm{s}^2$ continues to evolve and deviates from $\req$, although very slowly. This is because \cref{eq:general_req} for $\req$ depends on $\theta$, which increases with time. This increase in $\theta$ then leads to slow changes in $\req$ as time goes by. We also notice that $c_\mathrm{s}^2$ is always positive for $p=1.8 $, while it is negative at late times and in the future for $p=2.3$. Our numerical analysis shows that in both cases the semiequilibrium values of $c_\mathrm{s}^2$ are in excellent agreement with the values given by \cref{eq:c_s_s_p}. As for \cref{eq:c_s_s_p_2}, however, the values it provides agree with those given by \cref{eq:c_s_s_p} only for $p=2.3$ and not for $p=1.8$. As expected, this is due to the fact that $r_0/\req$ is not very small for the latter case, and therefore, contributes significantly to $c_\mathrm{s}^2$; we have confirmed this numerically. Finally, the observation that $c_\mathrm{s}^2$ becomes negative at late times for the $p=2.3$ case (and stays negative) implies that this specific example is plagued by gradient instabilities and therefore does not provide a theoretically viable cosmological evolution. In order to see how excluding solutions with gradient instabilities further constrains the parameter space of a given model, we add to the exclusion plots of \cref{fig:p_m_parameter} a region in light gray which corresponds to parameter combinations for which the sound speed squared becomes negative at least once during the cosmic history, i.e., before the present time. The upper panel of the figure shows that this happens only for $p>2$, which is in agreement with the approximate \cref{eq:c_s_s_p,eq:c_s_s_p_2}.
\begin{figure}
	\centering
	\includegraphics[width=1\linewidth]{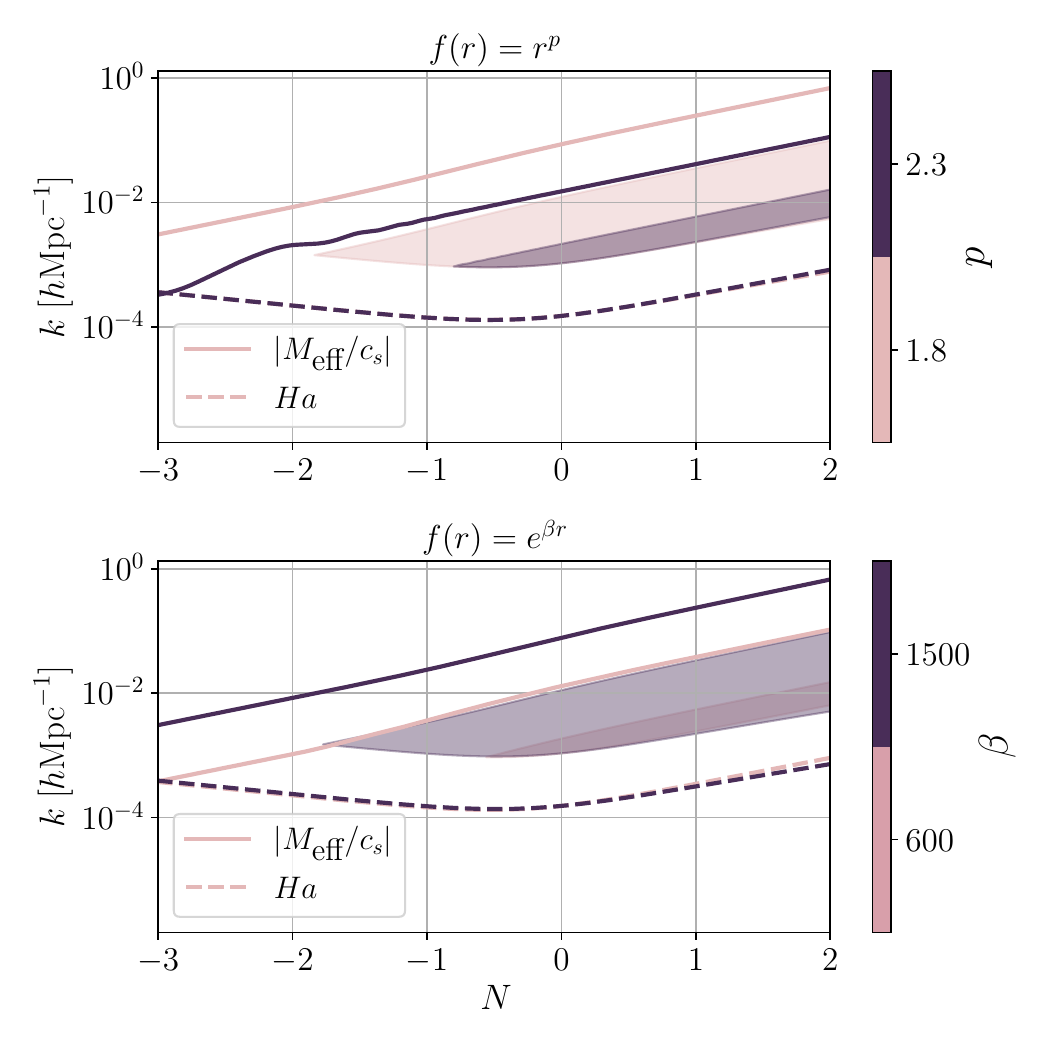}
	\caption{Ranges of scales $k$ for which the approximate analytic expression \eqref{eq:cs2_general} for the sound speed of the light mode of cosmological linear perturbations is valid, i.e., values of $k$ which satisfy the condition $H^2a^2 \ll k^2 \ll \Meff^2c_\mathrm{s}^{-2}$. These ranges are shown as functions of the number of $e$-folds $N$, where $N=0$ corresponds to the present time. The upper and lower panels correspond to power-law and hyperbolic metrics, respectively, and the  parameters of the models are set to the same values used in \cref{fig:cs}. These are marked with letters A, B, C, and D in \cref{fig:p_m_parameter,fig:m_beta_param}.}
	\label{fig:pert_k}
\end{figure}

For the hyperbolic field-space metric, we have
\begin{equation}
    \mathcal{R}=-\frac{1}{2}\beta^2\,,
\end{equation}
which results in
\begin{equation}
    \Meff^2 \approx a^2 m^2 (1-\beta \req)\,,
\end{equation}
where we have set $r_0=0$ and again assumed that the $r$ field is at its semiequilibrium value $\req$. Using this value for $\Meff^2$ and the value of $\Omega^2$ given by \cref{eq:turningrateexp}, \cref{eq:cs2_general} for the sound speed gives
\begin{equation}
    \label{eq:cs2_exp}
    c_\mathrm{s}^2 = \frac{1-\beta\req}{1+\beta\req}\,.
\end{equation}
As discussed in \cref{sec:hyp-metric}, $\beta \req$ can be smaller or larger (although not much larger) than unity. In the latter case of $\beta \req>1$, \cref{eq:cs2_exp} then implies that the hyperbolic metric leads to a negative $c_\mathrm{s}^2$ for the light mode once $r$ has taken its semiequilibrium value $\req$. We show in the lower panel of \cref{fig:cs} two examples of the time evolution of $c_\mathrm{s}^2$ for the hyperbolic metric. The $\beta=600$ case is an example of $\beta\req>1$ (as $\beta\req=2.35$) while $\beta\req<1$ for the $\beta=1500$ case (as $\beta\req=0.93$). The figure shows that for the smaller value of $\beta$ (i.e., $\beta=600$), $c_\mathrm{s}^2$ becomes negative at some point in the past and stays negative after that, which is consistent with the negative semiequilibrium value ($\sim-0.4$) given by \cref{eq:cs2_exp}. For the $\beta=1500$ case, however, the sound speed squared is positive for the entire cosmic history and approaches the (positive) semiequilibrium value ($\sim0.036$) at a point in the future. It is important to note that $c_\mathrm{s}^2$ continues to evolve after that point, deviating more and more from the value shown by the dashed horizontal line in the figure, and eventually becomes negative. That is because, as we discussed earlier for power-law metrics, the semiequilibrium value of $c_\mathrm{s}^2$ shown in the figure is based on the value of $\req$ computed at $\theta\sim0$, and since $\theta$ increases with time and \cref{eq:general_req} depends on $\theta$, the value of $\req$ also changes with time (although slowly for a large part of the parameter space) forcing the semiequilibrium value of $c_\mathrm{s}^2$ to change as well. The $\beta=1500$ case is therefore an example of cosmological solutions for which a negative sound speed squared occurs in the future and the light mode does not show any gradient instabilities over the cosmic history. Even though the existence of future gradient instabilities for a cosmological model may mean that it is theoretically unviable or at least less favored, we do not make a definite statement here on this and leave the interpretation of these results to the reader. Similar to the case of power-law metrics, we show in the exclusion plots of \cref{fig:m_beta_param} how the presence of gradient instabilities  (i.e., negative values of $c_\mathrm{s}^2$) in the past places further theoretical constraints on the parameter space of a given model with a hyperbolic field-space metric. This has been shown by the two light gray regions in both panels of \cref{fig:m_beta_param}.

Even though we discussed the constraints a negative $c_\mathrm{s}^2$ can place on our dark energy models and their parameters, it is important to remember that \cref{eq:cs2_general} and consequently \cref{eq:c_s_s_p,eq:cs2_exp} are approximate and valid only over the ranges of comoving scales given by the condition \eqref{eq:k_range}, which becomes
\begin{equation}
    H^2a^2 \ll k^2 \ll a^2m^2\left(p+2 - \frac{r_0}{\req}(p+1) \right)\label{eq:k_range_poly}
\end{equation}
for power-law metrics and
\begin{equation}
    H^2a^2 \ll k^2 \ll a^2m^2\left(1+\beta \req\right)\label{eq:k_range_hyper}
\end{equation}
for hyperbolic metrics. This means that if for a set of parameters no scales satisfy these conditions, then a negative $c_\mathrm{s}^2$ obtained from \cref{eq:cs2_general} does not exclude that parameter set. For the scales which do not satisfy the conditions, our approximate analytic expressions for the sound speed are not valid, and one then needs to solve the exact perturbation equations in order to numerically obtain the sound speed.  Given that the terms inside the parentheses on the right-hand sides of \eqref{eq:k_range_poly} and \eqref{eq:k_range_hyper} are of $\mathcal{O}(1)$, it is effectively the mass parameter $m$ that determines which modes \cref{eq:c_s_s_p,eq:cs2_exp} are valid for. The larger the value of $m$, the wider the range of scales for which \cref{eq:c_s_s_p,eq:cs2_exp} hold. In \cref{fig:pert_k}, we show ranges of scales $k$ as a function of $N$ which satisfy \eqref{eq:k_range} for the examples of \cref{fig:cs}. The solid and dashed curves in each case correspond, respectively, to the upper and lower bounds $|M_\mathrm{eff}/c_s|$ and $Ha$. The shaded regions then show the values of $k$ which satisfy \eqref{eq:k_range} at a given time, meaning that \cref{eq:cs2_general} is valid only for those values of $k$ at that time.\footnote{In drawing the validity regions of \cref{fig:pert_k}, we have assumed $50H^2a^2 < k^2 < 50\Meff^2c_\mathrm{s}^{-2}$.} We have marked the four cases of \cref{fig:cs,fig:pert_k} in the exclusion plots of \cref{fig:p_m_parameter,fig:m_beta_param} by letters A and B for the power-law metrics (corresponding to $p=2.3$ and $p=1.8$, respectively) and letters C and D for the hyperbolic metrics (corresponding to $\beta=600$ and $\beta=1500$, respectively). We have deliberately chosen one small and one large value for the parameter $m$ in each case, so that we illustrate the effect of $m$ on the width of the $k$ range. As both panels of \cref{fig:pert_k} demonstrates, the ranges of scales for the benchmark points B and D (with larger values of $m$) are wider compared to those for A and B with smaller values of $m$, as expected.

\section{Conclusions}
\label{sec:conclusions}

In this paper we have studied in detail cosmological dynamics of the models of multifield dark energy proposed in \rcite{Akrami:2020zfz}, where a number of scalar fields evolve along highly nongeodesic or ``spinning" trajectories in field space. We have particularly focused on models with two scalar fields, which we have called $r$ and $\theta$. We have assumed the field $r$ to be contributing to the potential energy through a simple quadratic function and a mass parameter of order of tens or hundreds of the Hubble constant. The other field, $\theta$, contributes to the potential through a linear function which breaks the $U(1)$ symmetry of the potential. We have additionally assumed that the field-space metric $\mathcal{G}_{ab}$ is diagonal with $\mathcal{G}_{rr}=1$ and $\mathcal{G}_{\theta\theta}=f(r)$, i.e., a function only of $r$. Contrary to the simple example presented in \rcite{Akrami:2020zfz}, where the field-space metric is assumed to be flat with $f(r)$ of quadratic form $r^2$, here we have allowed the function to be of a power-law form $r^p$ or a hyperbolic form $e^{\beta r}$, where $p$ and $\beta$ are arbitrary parameters.

Through an extensive phase-space analysis of the cosmological equations of motion at the background level, we have shown that not only do these classes of multifield dark energy models provide observationally viable cosmological dynamics, they are also theoretically appealing as the solutions satisfy a number of quantum-gravity-based conjectures (such as the de Sitter and distance swampland conjectures) for low-energy effective field theories describing the late-time universe. In particular, we have demonstrated that, for a well-motivated and large range of initial values of the fields and their derivatives, the models are able to provide accelerating solutions consistent with the present phase of the cosmic evolution, which are largely independent of the initial conditions. These solutions behave nearly as attractors even though the equations of motion in general do not form a closed autonomous dynamical system. We have illustrated this behavior through an extensive numerical analysis of models with different values of parameters for both the potential and the field-space metrics. We have also derived, for both power-law and hyperbolic metrics, a large number of approximate analytic equations, conditions, and expressions for different quantities, and demonstrated that they agree very well with the exact numerical results. We have proved, in particular, that for a wide range of models and parameters, and independently of initial conditions, the field $r$ quickly reaches a semiequilibrium value $\req$ which changes with time only slowly after that, allowing us to describe the entire cosmic history with a set of simple approximate equations. We have provided approximate (and simple) analytic forms for this semiequilibrium value $\req$, as well as the turning rate $\Omega$ and the dark energy equation of state $w_\phi$.

We have additionally investigated what constraints various theoretical and observational requirements would place on the models and their free parameters. We have done this through a combination of analytical and numerical analyses. We have shown, particularly, that the requirement of $w_\phi$ and $\Omega_\phi$ (the dark energy fractional density parameter) being close to $-1$ and $0.7$ today, respectively, highly constrains the parameter space of a given model. We have also demonstrated that the additional requirement that $\Omega \gg H$ at all times, with $H$ the Hubble expansion rate, which is a necessary condition for steep potentials to provide viable cosmic acceleration, further constrains the parameters. This latter condition is also required for the de Sitter swampland bounds to be satisfied. Our conclusion therefore is that large parts of the parameter space exist for both power-law and hyperbolic field-space metrics which provide cosmological solutions consistent with both observational viability requirements and theoretically desired conditions.

Finally, we have briefly analyzed linear cosmological perturbations in our multifield dark energy models by focusing on the light mode of the perturbations in the regime where the scales are (1) subhorizon and (2) larger than the Compton wavelength of the heavy mode. We have studied the sound speed of perturbations over this range of scales through both numerically solving the equations and approximating analytic expressions for the sound speed. We have computed these quantities for both power-law and hyperbolic metrics, and for different sets of parameters, and we have demonstrated that they all agree. We have shown that for certain choices of parameter values, the sound speed squared of the light mode becomes negative at some point in the past or in the future, which implies that the perturbations are plagued by gradient instabilities. We have shown how the requirement that such instabilities do not happen, at least over the cosmic history, further constrains the parameters of the models.

In summary, we have performed in this paper a detailed investigation of the dynamical properties of nongeodesic multifield dark energy models through a combination of analytical and numerical analyses and by studying, in detail, a simple but well-motivated and rich class of such models. We have shown how a combination of observational and theoretical requirements can easily be satisfied by the models while constraining their free parameters.

The next step in the analysis of these models is to investigate them and explore their parameters (1) beyond the background level and the approximations made about the cosmological perturbations, (2) by obtaining predictions of the models for different cosmological observables, and (3) by performing an extensive and rigorous statistical analysis of the models and exploring their parameter spaces by confronting them with various observational data. It is particularly expected, as discussed in this paper and shown in \rcite{Akrami:2020zfz}, that these models, in spite of being largely indistinguishable from the standard $\Lambda$CDM model at the background level, result in dark energy clustering on subhorizon scales and therefore lead to an enhanced growth of large-scale structure. This statistical analysis of the models where dark energy clustering is studied extensively with the objective of identifying potential observational signatures is the subject of an accompanying paper \cite{Eskilt2022}.

\acknowledgements{
JRE is supported by the European Research Council (ERC) under the Horizon 2020 Research and Innovation Programme (Grant agreement No. 819478). YA is supported by Richard S. Morrison Fellowship and LabEx ENS-ICFP: ANR-10-LABX-0010/ANR-10-IDEX-0001-02 PSL*. ARS's research is partially supported by funds from the Natural Sciences and Engineering Research Council (NSERC) of Canada. Research at the Perimeter Institute is supported in part by the Government of Canada through NSERC and by the Province of Ontario through MRI. VV is supported by the WPI Research Center Initiative, MEXT, Japan and partly by JSPS KAKENHI grants (20H04727, 20K22348).
}


\bibliography{refs}

\end{document}